\documentclass[12pt]{iopart}
\usepackage{epsfig}

\newcommand{\be}{\begin{equation}}
\newcommand{\ee}{\end{equation}}
\newcommand{\bea}{\begin{eqnarray}}
\newcommand{\eea}{\end{eqnarray}}
\newcommand{\beastar}{\begin{eqnarray*}}
\newcommand{\eeastar}{\end{eqnarray*}}
\newcommand{\nn}{\nonumber\\}

\newcommand{\Tc}{T_{\rm c}}
\newcommand{\tw}{t_{\rm w}}
\newcommand{\yw}{y_{\rm w}}

\newcommand{\order}{{{\mathcal O}}}
\newcommand{\ie}{{\it i.e.}}
\newcommand{\eg}{{\it e.g.}}

\newcommand{\half}{\frac{1}{2}}
\newcommand{\rv}{\mathbf{r}}

\newcommand{\om}{\omega}
\newcommand{\omt}{w}

\newcommand{\eq}[1]{~(\ref{#1})}

\renewcommand{\sc}[1]{{\mathcal{F}}_{#1}}

\newcommand{\zv}{\mathbf{0}}
\newcommand{\qv}{\mathbf{q}}
\newcommand{\dq}{\int(dq)\,}

\renewcommand{\sc}[1]{{\mathcal{F}}_{#1}}
\newcommand{\G}{{\mathcal{G}}}

\newcommand{\Ltwo}{L^{(2)}}
\renewcommand{\eql}{_{\rm eq}}

\newcommand{\CCtd}{\tilde\gamma_d}

\newcommand{\taum}{\tau_{\rm m}}
\begin{document}

\title[Fluctuation-dissipation relations in critical coarsening]{Fluctuation-dissipation relations in 
critical coarsening: crossover from
unmagnetized to magnetized initial states}

\author{Alessia Annibale\footnote[1]{Email
alessia.annibale@kcl.ac.uk}, Peter Sollich\footnote[2]{Email
peter.sollich@kcl.ac.uk}}

\address{King's College London, Department of Mathematics, London WC2R
2LS, UK}

\begin{abstract}
We study the non-equilibrium dynamics of the spherical ferromagnet
quenched to its critical temperature, as a function of the
magnetization of the initial state.  The two limits of unmagnetized
and fully magnetized initial conditions can be understood as
corresponding to times that are respectively much shorter and much
longer than a magnetization timescale, as in a recent field
theoretical analysis of the $n$-vector model.  We calculate exactly
the crossover functions interpolating between these two limits, for
the magnetization correlator and response and the resulting
fluctuation-dissipation ratio (FDR). For $d>4$ our results match those
obtained recently from a Gaussian field theory. For $d<4$,
non-Gaussian fuctuations arising from the spherical constraint need to
be accounted for. We extend our framework from the fully magnetized
case to achieve this, providing an exact solution for the relevant
integral kernel. The resulting crossover behaviour is very rich, with
the asymptotic FDR $X^\infty$ depending non-monotonically on the
scaled age of the system. This is traced back to non-monotonicities of
the two-time correlator, themselves the consequence of large
magnetization fluctuations on the crossover timescale.
We correct a trivial error in our earlier calculation for fully
magnetized initial states; the corrected FDR is {\em
consistent} with renormalization group expansions to first order in
$4-d$ for the longitudinal fluctuations of the $O(n)$ model in the
limit $n\to\infty$.
\end{abstract}

\section{Introduction}
The use of fluctuation-dissipation ratios (FDR) has proved very
fruitful in the last decade or so for quantifying the
non-equilibrium dynamics of glasses and other systems exhibiting
aging.
In the context of mean-field spin glass models with infinite-range
interactions, the FDR, commonly denoted $X$, has been used to
formulate a {\em generalized} fluctuation-dissipation theorem (FDT)
where $X$ is interpreted in terms of an effective temperature, $T_{\rm
  eff}=T/X$ for the slow, non-equilibrated modes of the
system~\cite{CugKurPel97}.  The properties of $X$ and $T_{\rm eff}$
have attracted much attention, based on the hope they might allow
a generalized statistical mechanical description for
a broad class of non-equilibrium
phenomena~\cite{CriRit03,JacBerGar06,LeoMaySolBerGar07}.

However, the generalized FDT can be shown to hold exactly only for
infinite-range models. A matter of recent intense interest has been
whether the appealing features of this mean-field scenario survive in
more realistic systems with finite-range
interactions~\cite{CriRit03}.  A class of systems that has
proved useful in this context is represented by ferromagnets quenched
from high temperature to the critical temperature $\Tc$ or below (see
e.g.\ Refs.~\cite{GodLuc00b,CalGam02c,CalGam02,MayBerGarSol03,CalGam04} and the recent
review~\cite{CalGam05}).  The non-equilibrium dynamics in these
systems is due to coarsening, \ie\ the growth of domains with the
equilibrium magnetization (for $T<\Tc$) or equilibrium correlation
structure (for $T=\Tc$), and slows down as domain sizes increase.
In an infinite system, equilibrium is never reached, leading to aging;
the age-dependence of two-time quantities has a simple physical
interpretation in terms of the growth of the domain
lengthscale~\cite{Bray94}. Coarsening systems therefore provide a
physically intuitive setting for the study of aging phenomena as
observed \eg \ in glasses, polymers and 
colloids. They are, of course, not completely generic; compared to
e.g.\ glasses they lack features such as thermal activation over
energetic or entropic barriers.

We focus in this paper on critical coarsening, \ie\ coarsening at
$\Tc$, where interesting connections to dynamical universality exist.
The FDR $X$ is determined from correlation and response functions
which, in aging systems, depend on two times: the age $\tw$ of the
system and a later measurement time $t$. In contrast to mean-field
spin glasses, where $X$ is constant within each ``time sector'' (\eg\
$t-\tw=\order(1)$ vs $t-\tw$ growing with $\tw$), in critical
coarsening the FDR is a smooth function of $t/\tw$. This makes the
interpretation of $T/X$ as an effective temperature less obvious.  To
eliminate the time-dependence one can consider the limit of times that
are both large and well-separated. This defines an {\em asymptotic
  FDR}
\be
X^\infty=\lim_{\tw\to\infty}\lim_{t\to\infty}X(t,\tw)
\ee
An important property of this quantity is that it should be {\em
  universal}~\cite{GodLuc00b,CalGam05} in the sense that its value is
the same for different systems falling into the same universality
class of critical non-equilibrium dynamics.  This makes a study of
$X^\infty$ interesting in its own right, even without an
interpretation in terms of effective temperatures.

An intriguing theoretical question which has been addressed recently
is whether different {\em initial conditions} can lead to different
universality classes of critical coarsening.  Due to the universality
of $X^{\infty}$, these can be uncovered by studying the effect that
different initial conditions have on the FDR.  Of particular
interest has been the effect of an initial magnetization on the
ensuing coarsening. For the Ising model in high dimension or with
long-range interactions~\cite{GarSolPagRit05}, one finds that
magnetized initial states do produce a different value of $X^\infty$.
This suggests a different dynamical universality class from
conventional coarsening from unmagnetized states, even though the
magnetization decays to zero at long times.  Further steps in
this direction were taken in our recent calculation of exact FDRs for
magnetized coarsening below the upper critical dimension in the
spherical model~\cite{AnnSol06}. The propagation of a trivial error
meant that the results were at variance with the renormalization group (RG)
result of Ref.~\cite{CalGam07} derived for the longitudinal
fluctuations in the $n\to\infty$ limit of the $O(n)$ model within an expansion around $d=4$.
We give the corrected results in this paper, and
these are consistent with the RG calculations
(see~\ref{ref:corrections}). This suggests that the equivalence
between the dynamics of the spherical model and the large-$n$ limit of
the $O(n)$ model extends beyond the regime of Gaussian
fluctuations, where it is trivial to establish.

Recently it was emphasized in the context of a field-theoretic
analysis~\cite{CalGamKrz06} that 
one should think of the nonzero initial magnetization as
introducing a new timescale in the system. The two limits of
unmagnetized and magnetized initial conditions can then be understood
as corresponding to times that are respectively much shorter and much
longer than this magnetization timescale, and one can in fact
interpolate between these two limits using a crossover function that
depends on times scaled by the magnetization timescale.  This
crossover function was calculated in~\cite{CalGamKrz06} in the
classical (Gaussian) regime, \ie\ above the upper critical dimension,
but so far there are no predictions for this function for lower
dimensions where the critical behaviour is governed by non-mean-field
exponents. We provide the first results of this kind in this work by
calculating the relevant crossover functions exactly for the spherical
model in $2<d<4$, for the correlator, response and FDR of the magnetization.


In Sec.~\ref{sec:crossover_setup} we recall the known crossover
behaviour of the magnetization (which is directly related to a
function $g(t)$) and the general relations encoding the consequences
of this for the magnetization correlation and response functions. As
in~\cite{AnnSol06} non-Gaussian spin fluctuations are important and
will be accounted for via the kernel $L$. Key to our analysis for the
more complicated functions $g(t)$ in our current scenario is an {\em
  exact} solution of the integral equation defining $L$ that applies
independently of the time regime. In Sec.~\ref{sec:crossover_dgt4} we
then evaluate the magnetization correlator and response for $d>4$. As
expected, we find here full agreement with the Gaussian field-theoretic
calculations~\cite{CalGamKrz06}. Sec.~\ref{sec:crossover_dlt4} deals with
the more interesting case $d<4$. Here the analysis is more
complicated but we can still derive exact results for the asymptotic
FDR $X^\infty$.  The relevant crossover functions display unexpected
non-monotonicities that, close to the lower critical dimension $d=2$,
turn into singularities at intermediate values of the scaled system
age. We study carefully the relevant scaling regimes for $d\to
2$, and investigate how they arise from the behaviour of the two-time
magnetization correlator. Our results are summarized in
Sec.~\ref{sec:crossover_summary}.

\section{Setup of calculation and exact solution for $\Ltwo$}
\label{sec:crossover_setup}

We start by recapitulating briefly the relevant elements of our
previous analysis of critical coarsening in the spherical
ferromagnet~\cite{AnnSol06}. The model consists of $N$ spins $S_i$ on
a $d$-dimensional cubic lattice, with sites $\rv_i$ and Hamiltonian $H
= \half \sum_{(ij)} (S_i-S_j)^2$~\cite{Stanley68}. The spins are
real-valued but subject to the spherical constraint $\sum_i S_i^2=N$.
Langevin dynamics leads to a simple equation of motion for the Fourier
components $S_\qv = \sum_i S_i \exp(-i\qv\cdot\rv_i)$ of the spins,
$\partial_t S_\qv = -(\omega_\qv+z(t))S_\qv + \xi_\qv$
where $\omega_\qv = 2\sum_{a=1}^d (1-\cos q_a)$ is abbreviated to
$\omega$ below and $\xi_\qv$ is independent Gaussian noise on each
wavevector $\qv=(q_1,\ldots,q_d)$, with
$\langle \xi_\qv(t)\xi_\qv^*(t')\rangle = 2NT\delta(t-t')$. The
Lagrange multiplier $z(t)$ 
enforces the spherical constraint; as explained in~\cite{AnnSol06}, it
is in reality not just a simple function of time but a dynamical variable
with fluctuations of $\order(N^{-1/2})$ that cause all the non-trivial
effects in the behaviour of global observables. In terms of the
function
%
$g(t) = \exp\left(2\int_{0}^t dt'\, z(t')\right)$
%
the Fourier mode response is
\be
R_{\qv}(t,\tw)=\sqrt{\frac{g(\tw)}{g(t)}}
e^{-\om(t-\tw)}=\frac{m(t)}{m(\tw)}e^{-\om(t-\tw)}
\label{crossover:Rq}
\ee
In the second equality we have used that the time-dependent
magnetization can be written as $m(t)=(1/N)\langle
S_{\zv}(t)\rangle=R_\zv(t,0)(1/N)\langle
S_{\zv}(0)\rangle=m_0/\sqrt{g(t)}$ with $m_0=
(1/N)\langle
S_{\zv}(0)\rangle$ the initial magnetization. The full,
unsubtracted two-time correlator $C_\qv(t,\tw)=(1/N)\langle
S_\qv(t)S_\qv^*(\tw)\rangle$ can be related to its equal-time value by
the response function,
\be
C_{\qv}(t,\tw) = R_{\qv}(t,\tw)C_{\qv}(\tw,\tw)
\label{Cq_twotime}
\ee
The relevant equal-time value is given by
\bea
C_\qv(t,t) &=& \frac{C_\qv(0,0)}{g(t)}e^{-2\omega t} + 2T \int_0^t dt'\,
\frac{g(t')}{g(t)} e^{-2\omega(t-t')}
\label{Cqtt}
\eea

The function $g(t)$ is determined from the spherical constraint, which
imposes $\dq C_\qv(t,t)=1$. Here and below we abbreviate $(dq) \equiv
d\qv/(2\pi)^d$, where the integral runs over the first Brillouin zone
of the hypercubic lattice, i.e.\ $\qv\in[-\pi,\pi]^d$. The resulting
integral equation for $g(t)$ is
\be
g(t)=\dq C_{\qv}(0,0)e^{-2\om t}+2T\int_0^t dt'\,g(t')f(t-t')
\label{gt_eq}
\ee
with $f(t)=\dq \,e^{-2\om t}$. Our first task will be to understand
how the solution of this crosses over between the magnetized and
unmagnetized cases.
In terms of the Laplace transform $\hat{g}(s)=\int_0^{\infty} 
dt\,g(t)e^{-st}$, equation\eq{gt_eq} reads 
\be
\hat{g}(s)=\frac{1}{1-2T\hat{f}(s)}\dq\frac{C_{\qv}(0,0)}{s+2\om}
\label{gs_spherical_c}
\ee
We take as the initial condition the standard
choice~\cite{RitDie95,CalGamKrz06} 
of a small magnetization $m_0$ but otherwise uncorrelated spin fluctuations.
The initial equal-time Fourier mode correlator
can then be written as
\be
C_{\qv}(0,0)=\delta_{\qv,\zv}Nm_0^2+(1-m_0^2)
\label{initial_time_corr}
\ee
This unsubtracted correlator is $\order(1)$ for $\qv\neq \zv$ but
$\order(N)$ for $\qv=\zv$. (For the fluctuation-dissipation behaviour
we will need to look at the connected correlator $\tilde{C}_{\qv}$,
which is discussed below.) Equation\eq{initial_time_corr} yields,
bearing in mind that the integral $(dq)$ is really a sum over the $N$
discrete wavevectors with weight $1/N$ each,
\be
\dq\frac{C_{\qv}(0,0)}{s+2\om}=\frac{m_0^2}{s}+(1-m_0^2)\hat{f}(s)
\ee
Using this in\eq{gs_spherical_c} one has at criticality, where 
$T=\Tc=[\dq 1/\om]^{-1}=[2\hat{f}(0)]^{-1}$, 
\be
\hat{g}(s)=\hat{K}\eql^{-1}(s)\frac{1}{s}
\left[\frac{m_0^2}{s}+(1-m_0^2)\hat{f}(s)\right]
\label{g_Laplace}
\ee
with
\be
\hat{K}\eql(s)=\Tc\dq\frac{1}{\om(2\om+s)}
\label{crossover:K_eq}
\ee
the Laplace transform of the equilibrium form\eq{Keql} of the kernel $K$
defined below. 
As before~\cite{AnnSol06} we want to look at the long-time limit of
$g(t)$, corresponding to small $s$ in\eq{g_Laplace}. In this regime
$\hat{K}\eql(s)$ is given for $d>4$ by $\hat{K}\eql(0)-\hat{K}\eql(s)=
a s^{(d-4)/2}$ and for $d<4$ by $\hat{K}\eql(s)=b s^{(d-4)/2}$, with
$a$ and $b$ some $d$-dependent constants. In the remaining square
bracket of\eq{g_Laplace} only the first term is present
for a fully magnetized initial state ($m_0=1$); conversely, only the
second survives for the unmagnetized case ($m_0=0$). To see the
crossover between these limits the two terms need to be of the same
order. Because we are interested in small $s$ and $\hat f(0)$ is
nonzero, this implies that $m_0^2$ and $s$ must be of the same order. 
We then find to leading order in these small quantities
\be
\fl \hat{g}(s)=\left\{
\begin{array}{ll}
{\displaystyle{\hat{K}\eql^{-1}(0)\left[\frac{m_0^2}{s^2}+\frac{
\hat{f}(0)}{s}
\right]}} 
& (d>4)
\\
{\displaystyle{\frac{s^{(4-d)/2}}{b}\left[\frac{m_0^2}{s^2}+\frac{
\hat{f}(0)}{s}\right]}} & (d<4)
\end{array}
\right.
\label{g_smalls}
\ee
or in the time domain
\be
\fl g(t)=\left\{
\begin{array}{ll}
\hat{K}\eql^{-1}(0)\left[m_0^2 t + 
\hat{f}(0)
\right] & (d>4)
\\
{\displaystyle{\frac{1}{b}\left[\frac{m_0^2}{\Gamma(d/2)}t^{(d-2)/2} + \frac{
\hat{f}(0)}{\Gamma((d-2)/2)}t^{(d-4)/2}\right]}} & (d<4)
\end{array}
\right.
\label{g_longtime}
\ee
One can combine these two expressions as
\bea 
g(t) &=& \frac{1}{\mu_d} t^{-\kappa} (m_0^2 t + c) =
\frac{c}{\mu_d}t^{-\kappa} \left(\frac{t}{\taum}+1\right)
\label{gt}
\eea
where we have defined 
\be
\kappa=\left\{
\begin{array}{ll}
\frac{4-d}{2} & (d<4)\\
0 & (d>4)
\end{array}
\right.
\hspace{1 cm}
\mu_d=\left\{
\begin{array}{ll}
b\,\Gamma(d/2) & (d<4)
\\
\hat{K}\eql(0) & (d>4)
\end{array}
\right.
\ee
and $c=(1-\kappa)
\hat{f}(0)$. In the second equality of\eq{gt} we have taken out the
factor of $c$ to identify the crossover timescale
\be
\taum = \frac{c}{m_0^2}
\ee
which as anticipated in the introduction depends on the initial
magnetization of the system. In the time domain, our statement of the
relevant long-time scaling $m_0^2\sim s$ can now be phrased as
follows: we will be considering the limit of large $t$,
$\tw$ and $\taum$ (corresponding to small $m_0$) at fixed time ratios
$u_{\rm w}=\tw/\taum$ and $u_t=t/\taum$. For ease of comparison with
the work of~\cite{CalGamKrz06} we will write simply $u\equiv u_{\rm w}$
and mostly work with $u$ and the time ratio $x=t/\tw=u_t/u$ instead of
$u$ and $u_t$. In terms of these variables one can write the function
$g(t)$ as
%
%
\be
g(t)
=\frac{c}{\mu_d}\, t^{-\kappa}(ux+1)
\label{gt_taum}
\ee
For the magnetization one then finds
\be
m(t)=\frac{m_0}{\sqrt{g(t)}}
=\sqrt{\frac{c}{\taum}\,\frac{\mu_d}{c}} \frac{t^{\kappa/2}}{\sqrt{ux+1}}
=\frac{\mu_d^{1/2}}{t^{\alpha/2}}
\sqrt{\frac{ux}{ux+1}}
\label{magn_taum}
\ee
with the exponent $\alpha$ defined as $\alpha=1-\kappa$ as
in~\cite{AnnSol06}. The last square root equals unity for long times
if the initial magnetization is kept finite and nonzero (so that $u\gg
1$). Otherwise it gives the well-known correction to the fully
magnetized result when the initial magnetization is small, \ie\ when
$t\sim \taum$~\cite{RitDie95}. In particular, for $u_t=ux\ll 1$, the
magnetization displays critical initial slip, increasing as $m(t)\sim
t^{\kappa/2}$, before crossing over to the $t^{-\alpha/2}$ decay
around $u_t=1$. Our analysis for the fully magnetized case
in~\cite{AnnSol06} is now recognized as relating to the limit
$t,\tw\gg \taum$, and accordingly all results in this paper should
reduce to the ones in \cite{AnnSol06} in the limit $u\to\infty$.
(Loosely speaking, one can think of this limit as corresponding to
$\taum\to 0$, \ie\ ``$m_0=\infty$''~\cite{CalGamKrz06}.) In the
opposite limit $t,\tw\ll\taum$ we should get back the results for the
unmagnetized case $m_0=0$. In terms of our scaling variables, this
limit corresponds to $u\to0$ at fixed $x$. Note that there is in
principle a third, ``mixed'' regime where the earlier time $\tw\ll
\taum$ but the later time $t\gg \taum$, \ie\ $u\ll 1$ and $ux\gg 1$.
We will see, however, that essentially no new behaviour arises here
and the crossover between the magnetized and unmagnetized cases, which
the analysis below will allow us to elucidate explicitly, is governed
principally by $u$. 

We next explore how the crossover effects in $g(t)$ modify the expressions for 
the long-time behaviour of the connected Fourier mode correlator 
$\tilde{C}_{\qv}(t,\tw)=C_\qv(t,\tw)-(1/N)\langle
S_\qv(t)\rangle\langle S_\qv^*(\tw)\rangle = C_\qv(t,\tw) -
N\delta_{\qv,\zv} m(t)m(\tw)$ and
the response function $R_{\qv}(t,\tw)$. 
(Here, as previously, we will not write explicitly the dependence on
$\taum$.) From~\cite{AnnSol06} we know that the equal-time 
connected correlator has the same expression as the unsubtracted
correlator
\be
\tilde{C}_{\qv}(\tw,\tw)=\frac{1}{g(\tw)}[\tilde{C}_{\qv}(0,0)e^{-2\om
  \tw}+2\Tc\int_0^{\tw} dt'\,e^{-2\om(\tw-t')}g(t')]
\label{longtime_Cq}
\ee
except for the appropriately modified initial condition
$\tilde{C}_{\qv}(0,0)=1-m_0^2$ which -- in contrast to the
unsubtracted $C_{\qv}$ -- is $\order(1)$ for all $\qv$.
For the zero Fourier mode one sees that in the long-time limit the
first term is subleading and the integral diverges at the upper end so
that one can use the asymptotics of $g(t')$, giving
\be
\fl \tilde{C}_{\zv}(\tw,\tw)=\frac{1}{g(\tw)}[\tilde{C}_{\zv}(0,0)+2\Tc\int_0^{\tw} dt'\,g(t')]
= 2\Tc\tw\frac{u/(2-\kappa)+1/(1-\kappa)}{u+1}
\label{longtime_Czero}
\ee
Similarly in the ratio of nonzero and zero mode correlators, expressed
in terms of the scaling variable $w=\om\tw$,
\bea
\fl\frac{\tilde{C}_{\qv}(\tw,\tw)}{\tilde{C}_{\zv}(\tw,\tw)}&=&\frac{(1-m_0^2)e^{-2w}+2\Tc\tw\int_0^1 dz\,e^{-2w(1-y)}g(z\tw)}{1+2\Tc\tw\int_0^1 dz\,g(z\tw)}
\eea
one can neglect the non-integral terms for long times and gets
\bea
\fl\frac{\tilde{C}_{\qv}(\tw,\tw)}{\tilde{C}_{\zv}(\tw,\tw)}&=&
\frac{\int_0^1 dz\,e^{-2w(1-y)}z^{-\kappa}(zu+1)}
{\int_0^1 dz\,z^{-\kappa}(uz+1)}
=\frac{\int_0^1 dz\,e^{-2w(1-z)}z^{-\kappa}(uz+1)}{u/(2-\kappa)+1/(1-\kappa)}
\label{nonzero_scaled}
\eea
Putting the last two results together yields the general scaling
\be
\fl \tilde{C}_{\qv}(\tw,\tw)=\frac{\Tc}{\om}\sc{C}(w,u), \,\,\,\,
\sc{C}(w,u)=\frac{2w}{u+1}\int_0^1 dz\,e^{-2w(1-z)}z^{-\kappa}(uz+1)
\label{scC_general_u}
\ee
One checks easily that $\sc{C}(w,u)$ reduces to the analogous scaling
functions for the unmagnetized and fully magnetized
cases~\cite{AnnSol06} in the appropriate limits $u\to 0$ and
$u\to\infty$. 
The magnetization response function is the zero mode response $R_\zv$.
From\eq{crossover:Rq}, using the scaling of the magnetization found
in\eq{magn_taum}, it is given by
\be
R_{\zv}(t,\tw)=\frac{m(t)}{m(\tw)}=x^{\kappa/2}\sqrt{\frac{u+1}{ux+1}}
\label{zero_R}
\ee

The results above are valid within the Gaussian approximation for the spin
dynamics in the spherical model, where the small fluctuations in the
Lagrange multiplier $z(t)$ are neglected.
As we saw in~\cite{AnnSol06}, in order to study the FD behaviour of the
magnetization (\ie\ of the zero Fourier mode, which is a global
observable) when an initial nonzero magnetization is present, we need
to account for non-Gaussian corrections arising from these Lagrange multiplier
fluctuations. Fortunately our earlier expressions~\cite{AnnSol06} for
the resulting magnetization correlator and response are valid for
arbitrary initial conditions and can be used directly. The
magnetization correlator including non-Gaussian effects is \cite{AnnSol06}
\be
C(t,\tw)= C^{(1)}(t,\tw)+C^{(2)}(t,\tw)
\ee
with
\bea
\fl C^{(1)}(t,\tw)&=&\tilde{C}_\zv(t,\tw)-\int dt'\,
    [M(t,t')\tilde{C}_\zv(\tw,t')+M(\tw,t')\tilde{C}_\zv(t,t')]m(t')
\nn
\fl & &{}+{}\int\, dt'\,d\tw'\, M(t,t')M(\tw,\tw')m(t')m(\tw')\tilde{C}_\zv(t',\tw')
\label{c_m1_unscaled}
\\
\fl &=& \int\, dt'\,d\tw'\, [\delta(t-t')-M(t,t')m(t')]
\nn
\fl & & \times [\delta(\tw-\tw')-M(\tw,\tw')m(\tw')]\tilde{C}_\zv(t',\tw')
\label{DM_part}
\eea
and
\be
C^{(2)}(t,\tw)=\half\int\, dt'\,d\tw'\, M(t,t')M(\tw,\tw')\tilde{C}\tilde{C}(t',\tw')
\label{c_m2}
\ee
where $\tilde{C}\tilde{C}(t',\tw')=\dq \tilde{C}_\qv^2(t',\tw')$. 
The corresponding expression for the global magnetization response including 
non-Gaussian effects is~\cite{AnnSol06} 
\bea
R(t,\tw)=
\int dt'\, [\delta(t-t')-M(t,t')m(t')]R_\zv(t',\tw)
\label{response_magn}
\eea
The key function $M$ appearing here is defined as follows. One starts
from the kernel
\be
K(t,\tw)=\dq R_{\qv}(t,\tw) C_{\qv}(t,\tw)=\dq R^2_{\qv}(t,\tw)
C_{\qv}(\tw,\tw)
\label{crossover:Kdef}
\ee
and its inverse $L$ defined by
\be
\int dt'\, K(t,t')L(t',\tw)=\delta(t-\tw)
\label{L_definition}
\ee
The behaviour of $K(t,\tw)$ near $\tw=t$ can be shown to imply the following
structure for $L$
\be
L(t,\tw)=\delta'(t-\tw)+2\Tc\delta(t-\tw)-L^{(2)}(t,\tw)
\label{L_structure_crossover}
\ee
where the first term arises from the fact that $K(t,\tw)$ is causal
(\ie\ it vanishes for $\tw>t$) and has a unit jump at $\tw=t$.
Finally, $M$ is defined to be proportional to the integral of $L$:
\be
M(t,\tw)=m(t)\int^t dt'\,L(t',\tw)
\label{crossover:Mdef}
\ee
In our previous analysis~\cite{AnnSol06} we had found long-time
scaling forms of $\Ltwo$ separately for the unmagnetized and
magnetized cases, with different methods needed for $d>4$ and $d<4$.
With the function $g(t)$ no longer being a simple power law, it seems
difficult if not impossible to adapt these methods to our
current crossover calculation. Fortunately, however, there is a
general and fully exact solution for $\Ltwo$ which applies in any
dimension and for any $g(t)$. To obtain this, we essentially integrate
by parts in\eq{L_definition}. In the derivative of $K$ with respect to
the earlier time argument we separate off the contribution from the
unit step and write
\be
\partial_{\tw} K(t,\tw) = -\delta(t-\tw) + K'(t,\tw)
\ee
where $K'$, like $K$, vanishes for $\tw>t$ and is finite elsewhere.
Correspondingly we split off the first term
from\eq{L_structure_crossover} and write
\be
\int^t dt'\,L(t',\tw) = \delta(t-\tw)+N(t,\tw)
\ee
where explicitly
\be
N(t,\tw)=2\Tc-\int_{\tw}^{t}dt'\, L^{(2)}(t',\tw)
\label{N_def}
\ee
and $N(t,\tw)$ also vanishes for $\tw>t$.
Integrating by parts in\eq{L_definition} and substituting these
definitions then yields
\be
K'(t,\tw)+\int_{\tw}^t dt'\, K'(t,t')N(t',\tw)-N(t,\tw)=0
\label{equation_for_N}
\ee
The point of this transformation is that non-equilibrium effects
manifest themselves in $K'$ in a very simple form. To see this, note
from\eq{Cqtt} for the unsubtracted correlator
that $\partial_{\tw}
C_{\qv}(\tw,\tw)=-[g'(\tw)/g(\tw)+2\om]C_{\qv}(\tw,\tw) + 2\Tc$, while
from\eq{crossover:Rq} $\partial_{\tw}
R_{\qv}^2(t,\tw)=[g'(\tw)/g(\tw)+2\om]R_{\qv}^2(t,\tw)$. Inserting
into\eq{crossover:Kdef} gives
\be
\fl K'(t,\tw)=2\Tc \dq R^2_{\qv}(t,\tw) = \frac{g(\tw)}{g(t)}\,2\Tc\dq
e^{-2\om(t-\tw)} = -\frac{g(\tw)}{g(t)} K\eql'(t-\tw)
\label{dK_general_sol}
\ee
where
\be
K\eql(t-\tw)=\dq \frac{\Tc}{\om}e^{-2\om(t-\tw)}
\label{Keql}
\ee
(with Laplace transform given by\eq{crossover:K_eq}) is the equilibrium form of
$K(t,\tw)$. With the simple multiplicative structure
of\eq{dK_general_sol} one can now solve the integral 
equation\eq{equation_for_N} for $N$ by inspection:
\be
N(t,\tw)=N\eql(t-\tw)\frac{g(\tw)}{g(t)}
\label{N_general}
\ee
where $N\eql(t-\tw)$ is the solution of the equilibrium version
of\eq{equation_for_N}, which is related to the corresponding
$\Ltwo\eql$ by
\be
\fl N\eql(t-\tw)=2\Tc-\int_0^{t-\tw}d\tau\, L\eql^{(2)}(\tau)
\approx\left\{
\begin{array}{ll}
\frac{2\lambda_d}{4-d}(t-\tw)^{(d-4)/2} & (d<4)
\\
\frac{1}{\mu_d}+\frac{2\lambda_d}{d-4}\, (t-\tw)^{(4-d)/2} & (d>4)
\end{array}
\right.
\label{Neql}
\ee
The last approximation gives the scalings for large time
differences $t-\tw$, derived from the corresponding asymptotic
behaviour of $\Ltwo\eql$. The latter is
$\Ltwo\eql(t-\tw)=\lambda_d(t-\tw)^{(d-6)/2}$ in $d<4$ and
$\Ltwo\eql(t-\tw)=\lambda_d(t-\tw)^{(2-d)/2}$ in $d>4$, with
$\lambda_d$ a $d$-dependent coefficient~\cite{AnnSol06}. This
behaviour can be derived from the Laplace transform of $\Ltwo\eql$, 
which from the equilibrium versions of
(\ref{L_definition},\ref{L_structure_crossover}) follows as
\be
\hat L^{(2)}\eql(s) = s + \Tc - 1/\hat K\eql(s)
\label{LeqLT}
\ee
Note that $N\eql$ decays to
zero for $d<4$ because $\hat L^{(2)}\eql(0) = \int_0^\infty d\tau
\Ltwo\eql(\tau)=2\Tc$ exactly, while for $d>4$ it approaches the
nonzero limit $2\Tc-\hat L^{(2)}\eql(0) = 1/\mu_d$~\cite{AnnSol06}.

The kernel $M$ is directly related to $N$ from\eq{crossover:Mdef}
and\eq{N_def}:
\be
M(t,\tw)=m(t)[\delta(t-\tw)+N(t,\tw)]
\label{M_from_N}
\ee
and in our current context we do not then need to compute $\Ltwo$
explicitly. Briefly, though, the general solution for $\Ltwo$ is
\bea
\fl L^{(2)}(t,\tw)&=&-\partial_t N(t,\tw)=
\frac{g(\tw)}{g(t)}\Ltwo\eql(t-\tw)+\frac{g'(t)g(\tw)}{g^2(t)}N\eql(t-\tw)
\label{Ltwo_cross}
\eea
and we outline in~\ref{sec:Ltwo_comparison} how this retrieves all of
our previous results in the appropriate limits. 
The key advantage of the above solution method is that it
automatically accounts for all non-equilibrium effects by reducing the
problem to an equilibrium calculation at criticality, where all
functions depend only on time differences and the relevant integral
equation can easily be solved by Laplace transform as shown in\eq{LeqLT} above.

With the general solution for $M(t,\tw)$, and hence for the magnetization
correlator and response, now in hand we analyse separately the cases 
$d>4$ and $d<4$.

\section{Crossover behaviour in $d>4$}
\label{sec:crossover_dgt4}

We first consider the situation $d>4$ above the upper critical
dimension. We expect to find here the same results for universal
quantities as in the Gaussian field theory of~\cite{CalGamKrz06}.
The zero Fourier mode Gaussian correlator and response are obtained
from\eq{longtime_Czero} and\eq{zero_R} by setting $\kappa=0$:
\be
R_{\zv}(t,\tw)=\sqrt{\frac{u+1}{ux+1}}\ ,\qquad
\tilde{C}_{\zv}(\tw,\tw)=\Tc\,\tw\frac{u+2}{u+1}
\ee
With these, one has from\eq{N_general},\eq{Neql} and\eq{M_from_N}
\bea
\fl M(t,t')m(t')&=&\frac{m_0^2}{\sqrt{g(t)g(t')}}\left\{\delta(t-t')+\frac{g(t')}{g(t)}\left[\frac{1}{\mu_d}+\frac{2\lambda_d}{d-4}(t-t')^{(4-d)/2}\right]\right\}\nn
\fl &=&\frac{\mu_d u}{\tw}\frac{1}{(ux+1)^{1/2}(uy+1)^{1/2}}
\Biggl\{
\tw^{-1}\delta(x-y)
\nn
\fl & &{}+{}\frac{uy+1}{ux+1}\left[\frac{1}{\mu_d}+\frac{2\lambda_d}{d-4}\tw^{(4-d)/2}(x-y)^{(4-d)/2}
\right]
\Biggr\}
\eea
where we have rescaled the times with $\tw$ and introduced the scaling 
variable $y=t'/\tw$.
In the long-time limit the first and the third terms in the above expression 
are subleading for $d>4$ so
\be
M(t,t')m(t')=\frac{u(uy+1)^{1/2}}{\tw(ux+1)^{3/2}}
\ee
By inserting this expression
into\eq{response_magn} one finds the magnetization response
\bea
R(t,\tw)&=&R_\zv(t,\tw)- \int dt'\, M(t,t')m(t')R_\zv(t',\tw)
\nn
&=&\left(\frac{u+1}{ux+1}\right)^{1/2}-\frac{u(u+1)^{1/2}}{(ux+1)^{3/2}}(x-1)
\ =\ \left(\frac{u+1}{ux+1}\right)^{3/2}
\eea
The magnetization correlator is found from\eq{c_m1_unscaled} and reads after
rescaling all times 
\bea
\fl C^{(1)}(t,\tw)&=&\Tc \tw
\Biggl\{
\frac{u+2}{u+1}\left[\sqrt{\frac{u+1}{ux+1}}-\int_1^x dy\,\frac{u(u+1)^{1/2}}{(ux+1)^{3/2}}\right]-
\int_0^1 dy\, \frac{uy(uy+2)}{(ux+1)^{3/2}(u+1)^{1/2}}\nn
\fl
& & {}-{}\left.\int_0^1 dy\,\frac{uy(uy+2)}{(u+1)^{3/2}(ux+1)^{1/2}}
+\int_1^x dy\,\int_0^1 d\yw\,\frac{u^2\yw(u\yw+2)}{(ux+1)^{3/2}(u+1)^{3/2}}
\right.
\nn
\fl & &
{}+{}\left.\int_0^1 dy\,\int_0^yd\yw\,\frac{u^2\yw(u\yw+2)}{(ux+1)^{3/2}(u+1)^{3/2}}
\right.
\nn
\fl & &{}+{}\int_0^1 dy\,\int_y^1 d\yw\,\frac{u^2y(uy+2)}{(ux+1)^{3/2}(u+1)^{3/2}}
\Biggr\}
\nn
\fl &=&
\Tc\tw\frac{2+3u+2u^2+\frac{1}{2}u^3}{(u+1)^{3/2}(ux+1)^{3/2}}
\label{Cm_dabove4}
\eea
The term $C^{(2)}$ scales as $\sim\tw^{(4-d)/2}$ for $4<d<6$, where the 
integral\eq{c_m2} that defines it can be shown to be dominated by aging 
timescales, and as $\sim\tw^{-1}$ for $d>6$, where
$\tilde{C}\tilde{C}(t',\tw')$ in\eq{c_m2} 
behaves as a short range kernel, so it is always subleading. 
Thus\eq{Cm_dabove4} represents the full long-time magnetization
correlator for $d>4$.

The $t$-dependence in the correlator $C\equiv C^{(1)}$ is the same as in the response
$R$ and only occurs via the overall factor $(ux+1)^{-3/2}=(u_t+1)^{-3/2}$.
It therefore cancels in the resulting FDR which follows after a few lines
(using $\partial_{\tw}[\tw F(x,u)]=(1+u\,\partial_u-x\,\partial_x)F(x,u)$
to calculate $\partial_{\tw} C$) as
\bea
X(t,\tw)=\frac{\Tc R(t,\tw)}{\partial_{\tw}C(t,\tw)}
=\frac{4}{5}\frac{(u+1)^4}{(u+1)^4+\frac{3}{5}} \equiv X^\infty(u)
\label{FDR_above}
\eea
Thus, for $d>4$ the FDR is $t$-independent and hence identical to the
asymptotic FDR $X^\infty(u)=\lim_{t\gg\tw=u\taum\gg 1}X(t,\tw)$. It
interpolates between $1/2$ (for $u\ll 1$) and $4/5$ (for $u\gg 1$),
reproducing in these limits our previous results for the FDRs for
unmagnetized and fully magnetized initial conditions~\cite{AnnSol06}.
As expected from the universality of $X^\infty$, our result for the
entire crossover function also exactly agrees with that calculated from a
Gaussian field theory~\cite{CalGamKrz06}. 
\begin{figure}
\setlength{\unitlength}{0.40mm} 
\begin{picture}(200,155)(-100,10)
\put(10,25){\includegraphics[width=180\unitlength]{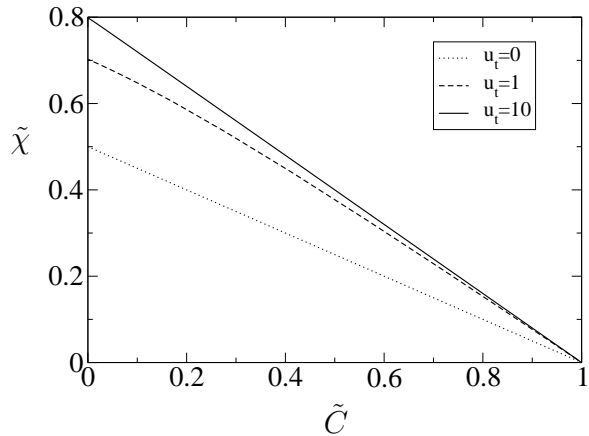}}
\put(-2,105){$\tilde{\chi}$}
\put(102,10){$\tilde{C}$}
\end{picture}
\caption{Normalized magnetization FD plot for dimensionality $d$ above $4$, 
showing the normalized susceptibility 
$\tilde{\chi}$ versus the normalized correlation $\tilde{C}$, for different 
fixed values of $u_t=u x$ as indicated in the figure. For $u_t=0$ the
plot is a straight  
line with (negative) slope $1/2$, as expected from the unmagnetized limit. 
As $u_t$ is increased the initial slope of the plot converges  
quickly to $4/5$, corresponding to the fully magnetized limit, 
and the crossover to the unmagnetized regime occurs at larger time 
differences and eventually becomes invisible on the scale of the
plot. } 
\label{fig:FDabove4}
\end{figure}
The FD plot is obtained by graphing the normalized 
susceptibility $\tilde\chi(t,\tw)=\Tc\chi(t,\tw)/C(t,t)$ versus the
normalized correlator $\tilde C(t,\tw)=C(t,\tw)/C(t,t)$
at fixed $u_t=u x$ and using $x$ (or $u$) as the curve parameter. The
factor of $\Tc$ is included in the definition of $\tilde\chi$ to make
the equilibrium FD plot a line of (negative) slope 1.
The susceptibility is obtained from $R$ by integration as usual,
\be
\fl \chi(t,\tw)=\int_{\tw}^t dt'\,R(t,t')=
\frac{\tw}{u}\int_{u}^{u_t} du'\,R(u_t,u') = 
\frac{2}{5}\frac{\tw}{u}\frac{(ux+1)^{5/2}-(u+1)^{5/2}}{(ux+1)^{3/2}}
\label{cross_chi_dgt4}
\ee
(with some obvious abuse of notation in the representation as an integral over $u'$). 
The results, displayed in Fig.~\ref{fig:FDabove4}, show that for $u_t=0$,
the curve is a straight line with (negative) slope $1/2$, 
as expected from the unmagnetized limit.
As $u_t$ is increased, the initial slope of the plot converges 
quickly to $4/5$, which is expected in the fully magnetized regime, 
and the unmagnetized regime gets progressively squeezed into the top 
left corner of the plot where it eventually becomes
invisible. Intuitively, this is because for $u_t=t/\taum\gg1$ we need
to move to 
relatively much earlier times
$u=\tw/\taum \sim 1$ in order for the dynamics to be sensitive to the fact
that the initial magnetization was small.

\section{Crossover behaviour in $d<4$}
\label{sec:crossover_dlt4}

In $d<4$ the analysis is somewhat more awkward and leads to highly non-trivial 
magnetization FD behaviour as we will see. As before we will find that
all relevant quantities vary on aging timescales $\sim\tw$ and so
we will exploit the relevant asymptotics for large time
differences throughout.

One starts by working out the combination $M(t,\tw)m(\tw)$ appearing in the 
definition of $C$ and $R$: 
\bea
\fl M(t,\tw)m(\tw)&=&N(t,\tw)m(\tw)m(t)=
\frac{m_0^2}{\sqrt{g(t)g(\tw)}}\frac{g(\tw)}{g(t)}
\frac{2\lambda_d}{4-d}(t-\tw)^{(d-4)/2}
\label{M_initial}
\\
\fl &=&\frac{2\lambda_d \mu_d m_0^2}{(4-d)c}\frac{\tw^{-3\kappa/2}}
{t^{-3\kappa/2}}\frac{(u+1)^{1/2}}{(ux+1)^{3/2}}(x-1)^{(d-4)/2}
\ =\ \frac{1}{t}\sc{M}(x,u)
\label{scaling_M}
\eea
where we have defined
\be
\sc{M}(x,u)=\frac{d-2}{2}u x^{(8-d)/4}\frac{(u+1)^{1/2}}{(ux+1)^{3/2}}\left(\frac{x-1}{x}\right)^{(d-4)/2}
\label{sc_M}
\ee
and used $2\lambda_d \mu_d=(4-d)(d-2)/2$~\cite{AnnSol06}.
Equation\eq{sc_M} represents the generalization to finite $u$ of the 
scaling function $\sc{M}(x)$ determined in \cite{AnnSol06} and reduces to the
latter in the limit $u\rightarrow\infty$ as it should. Note that
in\eq{M_initial} we have directly neglected the contribution
$\delta(t-\tw)m(\tw)m(t)\sim
\delta(x-1)\tw^{-1}\tw^{-\alpha/2}t^{-\alpha/2} \sim
\tw^{\kappa-2}=\tw^{-d/2}$ because it is subleading for long times
compared to the main $1/t\sim 1/\tw$ term in\eq{scaling_M}. 

We note briefly the explicit expression
\be
\sc{M}\!\left(\textstyle\frac{x}{y},uy\right)=
\frac{d-2}{2}u\, x^{4-3d/4} 
\frac{(uy+1)^{1/2}}{(ux+1)^{3/2}} y^{(d-4)/4} (x-y)^{(d-4)/2}
\label{scM_xyuy}
\ee
that recurs in a number of calculations below; $\sc{M}(1/\yw,u\yw)$ is
obtained from this by replacing $x\to 1$, $y\to\yw$. 
It will also be useful for later to have the asymptotics of $\sc{M}(x,u)$
for large $x$,  
which will give the asymptotic behaviour of the correlator and thus of
$X^{\infty}$:
\be
\sc{M}(x,u)=\frac{d-2}{2} x^{(2-d)/4}\left(\frac{u+1}{u}\right)^{1/2}
\label{M_asymptotics}
\ee

For the response function the Gaussian contribution is
from\eq{zero_R}, after setting $\kappa=(4-d)/2$,
\be
R_\zv(t,\tw) = x^{(4-d)/4}\sqrt{\frac{u+1}{ux+1}}
\label{Rzero_Gauss}
\ee
The overall magnetization response is then found simply by inserting\eq{sc_M} 
into\eq{response_magn} and rescaling the times as before
\bea
\fl R(t,\tw) 
&=& x^{(4-d)/4}\sqrt{\frac{u+1}{ux+1}}
-\frac{d-2}{2}x^{3(4-d)/4}\frac{u(u+1)^{1/2}}{(ux+1)^{3/2}}\int_1^x dy\,(x-y)^{(d-4)/2}
\\
\fl &=& x^{(4-d)/4}\sqrt{\frac{u+1}{ux+1}}\left[1-\frac{ux}{ux+1}
\left(\frac{x-1}{x}\right)^{(d-2)/2}\right]
\label{R_generic_x}
\eea
For $u\gg 1$, one retrieves the fully magnetized limit 
calculated  previously~\cite{AnnSol06}. On the other hand, the
unmagnetized limit, $u\ll 1$, 
coincides with the Gaussian response. This is consistent
with the fact that non-Gaussian effects in the FD behaviour of the
(global) magnetization only need to be accounted 
for in the case of an initial nonzero magnetization~\cite{AnnSol06}.
The large-$x$ behaviour of $R(t,\tw)$ for general $u$, which will
provide the asymptotic FDR, is easily extracted from\eq{R_generic_x} as
\be
\fl R(t,\tw)= x^{-(d+2)/4}\left(\frac{d-2}{2}+\frac{1}{u}\right)\left(\frac{u+1}{u}\right)^{1/2}
\label{asymptotic_R}
\ee

We next turn to the correlator. 
Rescaling the times with $\tw$ in equation\eq{DM_part} one has for the
first part
\bea
\fl C^{(1)}(t,\tw)&=& \int_0^x\, dy\,\int_0^1d\yw\,
\left[\delta(x-y)-\frac{1}{x}\sc{M}\!\left(\textstyle
\frac{x}{y},uy\right)\right]
\nn
\fl & & \times \left[\delta(1-\yw)-\sc{M}\!\left(\textstyle
\frac{1}{\yw},u\yw\right)\right]\tilde{C}_\zv(\tw y,\tw \yw)
\label{DM_part_resc}
\eea
From\eq{Cq_twotime},\eq{longtime_Czero} and\eq{Rzero_Gauss}
the Gaussian factor can for long times be written as
\bea
\tilde{C}_\zv(\tw y,\tw\yw)&=&
2\Tc\tw [f(y,\yw)\theta(y-\yw)+f(\yw,y)\theta(\yw-y)]
\\
&=&
2\Tc\tw [f(y,\yw) + \theta(\yw-y)(f(\yw,y)-f(y,\yw))]
\label{C_subtraction}
\eea
where we have explicitly accounted for the ordering of the time arguments;
the dependence on $y$ and $\yw$ is through the function
\bea
f(y,\yw) &=&\left(\frac{y}{\yw}\right)^{\kappa/2} \yw \, 
\frac{u\yw/(2-\kappa)+1/(1-\kappa)}{u\yw+1}\,
\sqrt{\frac{u\yw+1}{uy+1}}
\\
&=&
y^{(4-d)/4}\yw^{d/4}\,
\frac{(2u\yw)/d+2/(d-2)}{(u\yw+1)^{1/2}(uy+1)^{1/2}}
\eea
Inserting\eq{C_subtraction} into\eq{DM_part_resc},
one can rewrite $C^{(1)}$ as 
\bea
\fl \frac{C^{(1)}(t,\tw)}{2\Tc\tw} &=& \int_0^x\, dy\,\int_0^1d\yw\, 
\left[\delta(x-y)-\frac{1}{x}
\sc{M}\!\left(\textstyle\frac{x}{y},uy\right)\right]
\left[\delta(1-\yw)-
\sc{M}\!\left(\textstyle\frac{1}{\yw},u\yw\right)\right]
\nn
\fl & &\times f(y,\yw)
\nn
\fl & & {}-{}
\int_0^1\, d\yw\,\int_0^{\yw}dy\,\frac{1}{x}
\sc{M}\!\left(\textstyle\frac{x}{y},uy\right)
\left[\delta(1-\yw)-\sc{M}\!\left(\textstyle\frac{1}{\yw},u\yw\right)\right]
\nn
\fl
& &\times[f(\yw,y)-f(y,\yw )]
\label{cancellation_analogue}
\eea
The decomposition\eq{C_subtraction} is the analogue of the
cancellation trick used for the fully magnetized  
case~\cite{AnnSol06}. There the analogue of the first line
in\eq{cancellation_analogue} 
vanished identically. This is not the case here, but the procedure
remains useful because it makes it easier to extract the large-$x$
limit: the first integral (denoted $F$ below) factorizes, and in the
second one (denoted $S$) both
integration variables $y,\yw$ are $\leq 1$ and so $\ll x$ for large $x$.
Using the factorization, the first double integral can be worked out
explicitly for generic $x$:
\bea
\fl F&=& \int_0^x\, dy\, \left[\delta(x-y)-\frac{1}{x}\sc{M}\!\left(\textstyle\frac{x}{y},uy\right)\right] \frac{y^{(4-d)/4}}{(uy+1)^{1/2}}
\\
\fl & & \times
\int_0^1d\yw\, 
\left[\delta(1-\yw)-\sc{M}\!\left(\textstyle\frac{1}{\yw},u\yw\right)\right]\yw^{d/4}\frac{(2u\yw)/d+2/(d-2)}{(u\yw+1)^{1/2}}
\label{F1}\nn
\fl
&=&\frac{x^{(4-d)/4}}{(ux+1)^{1/2}}
\left[
1-\frac{d-2}{2}x^{(4-d)/2}\frac{u}{ux+1}\int_0^xdy\,(x-y)^{(d-4)/2}
\right]
\nn
\fl & & \times
\int_0^1d\yw\, 
\left[\delta(1-\yw)-\sc{M}\!\left(\textstyle\frac{1}{\yw},u\yw\right)\right]\yw^{d/4}\frac{(2u\yw)/d+2/(d-2)}{(u\yw+1)^{1/2}}
\\
\fl &=& \frac{x^{(4-d)/4}}{(ux+1)^{3/2}}
\left[\frac{2u/d+2/(d-2)}{(u+1)^{1/2}}\right.
\nn
\fl
& & {}-{}\left.
\frac{d-2}{2}\frac{u}{(u+1)^{3/2}}
\int_0^1 d\yw\,\yw^{(d-2)/2}\left(\frac{2u\yw}{d}+\frac{2}{d-2}\right)(1-\yw)^{(d-4)/2}\right]
\eea
The remaining integral produces Beta functions so that 
\bea
\fl F &=& \frac{x^{(4-d)/4}}{(ux+1)^{3/2}}
\left\{\frac{2u/d+2/(d-2)}{(u+1)^{1/2}}
-\frac{u}{(u+1)^{3/2}}
\frac{\Gamma^2(d/2)}{\Gamma(d)}\left[u+\frac{2(d-1)}{d-2}\right]
\right\}
\label{F}
\eea
The large-$x$ behaviour is obtained by replacing the prefactor with
$x^{-(2+d)/4}/u^{3/2}$.
The second double integral in\eq{cancellation_analogue} can be written
explicitly as 
\bea
\fl S&=&{}-{}\frac{1}{x}\int_0^1 dy\,\sc{M}\!\left(\textstyle\frac{x}{y},uy\right)\frac{1}{(uy+1)^{1/2}(u+1)^{1/2}}
\nn
\fl & &\times \left[
y^{d/4}\left(\frac{2uy}{d}+\frac{2}{d-2}\right)
-y^{(4-d)/4}\left(\frac{2u}{d}+\frac{2}{d-2}\right)
\right] 
\nn
\fl & &{}+{}\frac{1}{x}\int_0^1 d\yw\,
\sc{M}\!\left(\textstyle\frac{1}{\yw},u\yw\right) \frac{\yw^{(4-d)/4}}{(u\yw+1)^{1/2}}
\nn
\fl & &\times\int_0^{\yw}\, dy\,\sc{M}\!\left(\textstyle\frac{x}{y},uy\right)
\frac{y^{d/4}}{(uy+1)^{1/2}}\left(\frac{2uy}{d}+\frac{2}{d-2}\right)
\nn
\fl & & {}-{}\frac{1}{x}\int_0^1 d\yw\,
\sc{M}\!\left(\textstyle\frac{1}{\yw},u\yw\right) \frac{\yw^{d/4}}{(u\yw+1)^{1/2}}\left(\frac{2u\yw}{d}+\frac{2}{d-2}\right)
\nn
\fl & &\times\int_0^{\yw}\, dy\,\sc{M}\!\left(\textstyle\frac{x}{y},uy\right)
\frac{y^{(4-d)/4}}{(uy+1)^{1/2}}
\\
\fl &=&
-\frac{d-2}{2}\frac{x^{3-3d/4}u}{(u+1)^{1/2}(ux+1)^{3/2}}
\int_0^1 dy\,(x-y)^{(d-4)/2} y^{(d-4)/4}
\nn
\fl & & \times\left[y^{d/4}\left(\frac{2uy}{d}+\frac{2}{d-2}\right)
-y^{(4-d)/4}\left(\frac{2u}{d}+\frac{2}{d-2}\right)\right]
\nn
\fl & & {}+{}\left(\frac{d-2}{2}\right)^2 \frac{x^{3-3d/4} u^2}{(u+1)^{3/2}(ux+1)^{3/2}}
\nn
\fl & &\times
\left[
\int_0^1d\yw\,(1-\yw)^{(d-4)/2}\int_0^{\yw}dy\,(x-y)^{(d-4)/2}y^{(d-2)/2}\left(\frac{2uy}{d}+\frac{2}{d-2}\right)
\right.
\nn
\fl & &
{}-{}\left.\int_0^1d\yw\,(1-\yw)^{(d-4)/2}\yw^{(d-2)/2}
\left(\frac{2u\yw}{d}+\frac{2}{d-2}\right)\int_0^{\yw}dy\, (x-y)^{(d-4)/2}
\right]
\eea
If we can now take the large-$x$ limit, where $(x-y)^{(d-4)/2}\approx
x^{(d-4)/2}$, the integrals can be carried out and the prefactor
simplifies, giving after a little algebra:
\bea
\fl
S &= &
\frac{x^{-(2+d)/4}}{u^{3/2}(u+1)^{3/2}}\left[
\frac{d-2}{d}u(u+1)\left(\frac{d}{d+2}u+1\right)
{}-
\frac{d-2}{2}
\frac{\Gamma^2(d/2)}{\Gamma(d)}u^2\left(\frac{u}{2}+1\right)
\right]
\label{S}
\eea
Gathering the contributions
from\eq{F} and\eq{S}, one gets for the large-$x$ limit of the first 
contribution to the correlator
\bea
\frac{C^{(1)}(t,\tw)}{2\Tc\tw}
&=&
\frac{x^{-(d+2)/4}}{u^{3/2}(u+1)^{3/2}}
\left(Au^3+Bu^2+Cu+D\right)
\label{asymptotic_c_m1}
\eea
with 
\bea
A&=&\frac{d-2}{d+2}-\frac{d-2}{4}\,\frac{\Gamma^2(d/2)}{\Gamma(d)}
\label{A}
\\
B&=&\frac{2d}{d+2}-\frac{d}{2}\,\frac{\Gamma^2(d/2)}{\Gamma(d)}
\label{B}
\\
C&=&\frac{d}{d-2}-\frac{2(d-1)}{d-2}\,
\frac{\Gamma^2(d/2)}{\Gamma(d)}
\label{C}
\\
D&=&\frac{2}{d-2}
\label{D}
\eea

The second contribution to the magnetization correlation comes from\eq{c_m2},
%
%
and to make progress here
we need the long time behaviour of the two-time function
$\tilde{C}\tilde{C}$ for the current case $d<4$. Proceeding as for the
fully magnetized scenario~\cite{AnnSol06}, we find first the scaling
of the equal-time value 
$\tilde{C}\tilde{C}(\tw,\tw)=\dq \tilde{C}_{\qv}^2(\tw,\tw)=\Tc^2\dq
\om^{-2}\sc{C}(\om\tw,u)$.
This is dominated by small $\om$, where
$(dq)=\sigma_d d\om\,\om^{(d-2)/2}$ with $\sigma_d$ the surface area
of a unit sphere in $d$ dimensions. Rescaling to $w=\om\tw$ gives
\be
\fl \tilde{C}\tilde{C}(\tw,\tw)=\gamma_d(u)\tw^{(4-d)/2}, \qquad
\gamma_d(u)=\Tc^2\sigma_d \int dw \,w^{(d-6)/2}\sc{C}^2(w,u) 
\ee
Normalizing $\tilde{C}\tilde{C}(t',\tw')$ with the equal time value $\tilde{C}\tilde{C}(t',t')$, one
obtains for $t'>\tw'$ in terms of the scaling variables $y=t'/\tw$ and
$\yw=\tw'/\tw$ (by rescaling in the numerator to $w=\om \tw'$ and in
the denominator to $w=\om t'$)
\bea
\fl \frac{\tilde{C}\tilde{C}(t',\tw')}{\tilde{C}\tilde{C}(t',t')}&=&\frac{g(\tw')}{g(t')}
\frac{\tw'^{(4-d)/2}}{t'^{(4-d)/2}}
\frac{\int dw\, w^{(d-6)/2}e^{-2w(y/\yw-1)}\sc{C}^2(w,u\yw)}
{\int dw\, w^{(d-6)/2}\sc{C}^2(w,uy)}\\
\fl &=&\frac{u\yw+1}{uy+1}\frac{\int dw\, w^{(d-6)/2}e^{-2w(y/\yw-1)}\sc{C}^2(w,u\yw)}{\int dw\, w^{(d-6)/2}\sc{C}^2(w,uy)}
\eea
For $t'<\tw'$, on the other hand, one has
\be
\fl \frac{\tilde{C}\tilde{C}(t',\tw')}{\tilde{C}\tilde{C}(t',t')}=\frac{\tilde{C}\tilde{C}(\tw',t')}{\tilde{C}\tilde{C}(\tw',\tw')}
\frac{\yw^{(4-d)/2}}{y^{(4-d)/2}}\frac{\gamma_d(u\yw)}{\gamma_d(uy)}
\ee
So overall
\be
\frac{\tilde{C}\tilde{C}(t',\tw')}{\tilde{C}\tilde{C}(t',t')} =\G\left(\frac{y}{\yw},u\yw\right)
\label{CC2_d_lt_4}
\ee
where
\be
\fl\G\!\left(\frac{y}{\yw},u\yw\right)=\left\{
\begin{array}{ll}
{\displaystyle\frac{u\yw+1}{uy+1}\frac{\int dw\, w^{(d-6)/2} \sc{C}^2(w,u\yw) e^{-2(y/\yw-1)w}}
{\int dw\, w^{(d-6)/2} \sc{C}^2(w,uy)}} & \mbox{for $y/\yw\geq 1$} \\
{\displaystyle\frac{\gamma_d(u\yw)}{\gamma_d(uy)} \left(\frac{y}{\yw}\right)^{(d-4)/2}\G(\yw/y,uy)} & \mbox{for $\yw/y\leq 1$}
\end{array}
\right.
\label{G_cross}
\ee
In the limit $u\gg 1$, this function should match with $\G(x)$ defined
in \cite{AnnSol06} for fully magnetized intial conditions, for
$x=y/\yw$. Unfortunately there was a typographical error in the
definition of $\G(x)$ as given in~\cite{AnnSol06}, which propagated
through the remainder of the calculation. In~\ref{ref:corrections} we
state the correct versions of all the relevant equations.  These
include, in particular, the first-order expansions around $d=4$ and
$d=2$ of $X^{\infty}$ in the fully magnetized case.

Having clarified the scaling behaviour of
$\tilde{C}\tilde{C}(t',\tw')$ in $d<4$, we can now work out $C^{(2)}$
from\eq{c_m2} by multiplying and dividing by $m(t')m(\tw')$ and
using\eq{scaling_M}
\bea
\fl C^{(2)}(t,\tw)&=&\half\int_0^t dt'\int_0^{\tw} d\tw'\, 
\frac{1}{m(t')m(\tw')\,t\,\tw}\sc{M}\!\left(\textstyle\frac{x}{y},uy\right)
\sc{M}\!\left(\textstyle\frac{1}{\yw},u\yw\right)
\nn
\fl &&\times
\tilde{C}\tilde{C}(t',t')\G\!\left(\textstyle\frac{y}{\yw},u\yw\right)
\\
\fl
&=&
\half\int_0^x dy \int_0^1 d\yw\,
\frac{\tw^{\alpha}}{x\mu_d}(y\yw)^{\alpha/2}
\frac{(uy+1)^{1/2}(u\yw+1)^{1/2}}{u(y\yw)^{1/2}}
\nn
\fl & &\times \sc{M}\!\left(\textstyle\frac{x}{y},uy\right)
\sc{M}\!\left(\textstyle\frac{1}{\yw},u\yw\right)
\tw^{(4-d)/2}y^{(4-d)/2}\gamma_d(uy)\G\!\left(\textstyle\frac{y}{\yw},u\yw\right)
\\
\fl
&=&\half\frac{\tw}{x u \mu_d} \int_0^x dy\,\int_0^1 d\yw\,
y^{(4-d)/4} \yw^{(d-4)/4} (uy+1)^{1/2}(u\yw+1)^{1/2}
\nn
\fl 
& & \times\sc{M}\!\left(\textstyle\frac{x}{y},uy\right)\sc{M}\!\left(\textstyle\frac{1}{\yw},u\yw\right)\gamma_d(uy)\G\!\left(\textstyle\frac{y}{\yw},u\yw\right)
\\
\fl
&=& \frac{\tw}{
x}\int_0^1 d\yw\,
\yw(u\yw+1)^{1/2}\sc{M}\!\left(\textstyle\frac{1}{\yw},u\yw\right)
\times
W
\label{cm2_sc_m}
\eea
where we have defined ($v=y/\yw$)
\be
\fl 2u\mu_d W=
\int_0^{x/\yw} dv\,v^{(4-d)/4}(uv\yw+1)^{1/2}\sc{M}\!\left(\textstyle\frac{x}{v\yw},uv\yw\right)
\gamma_d(uv\yw)\G\!\left(\textstyle v,u\yw\right)
\label{W}
\ee
To evaluate $W$ we split the integral into $v=0\ldots 1$ and
$v=1\ldots\infty$. In the former regime we rewrite $\G(v,\ldots)$ in terms of
$\G(1/v,\ldots)$ using\eq{G_cross} and then transform $v\to 1/v$ to
get
\bea
\fl 2u\mu_d W&=&
\int_0^1 dv\,v^{(d-4)/4}(uv\yw+1)^{1/2}\sc{M}\!\left(\textstyle\frac{x}{v\yw},uv\yw\right)
\gamma_d(u\yw)\G\!\left(\textstyle\frac{1}{v},uv\yw\right)
\nn
\fl 
& &{}+{}
\int_1^{x/\yw} dv\,v^{(4-d)/4}(uv\yw+1)^{1/2}\sc{M}\!\left(\textstyle\frac{x}{v\yw},uv\yw\right)
\gamma_d(uv\yw)\G\!\left(\textstyle v,u\yw\right)
\\
\fl
&=&
\int_1^{\infty} dv\,v^{-(d+4)/4}\left(\frac{u\yw}{v}+1\right)^{1/2}\sc{M}\!\left(\textstyle\frac{xv}{\yw},\frac{u\yw}{v}\right)
\gamma_d(u\yw)\G\!\left(\textstyle v,\frac{u\yw}{v}\right)
\nn
\fl 
& &{}+{}
\int_1^{x/\yw} dv\,v^{(4-d)/4}(uv\yw+1)^{1/2}\sc{M}\!\left(\textstyle\frac{x}{v\yw},uv\yw\right)
\gamma_d(uv\yw)\G\!\left(\textstyle v,u\yw\right)
\label{W_definition}
\eea
By performing the $w$-integrals in\eq{G_cross} explicitly, having
first inserted the definition\eq{scC_general_u} of $\sc{C}$, one finds for the terms
involving $\G$
\bea
\fl \frac{\gamma_d(u\yw)\G\!\left(\textstyle v,\frac{u\yw}{v}\right)}{\Tc^2\sigma_d}&=&
\frac{u\yw/v+1}{u\yw+1}\int dw\, w^{(d-6)/2}\sc{C}^2(w,u\yw/v)e^{-2w(v-1)}
\\
\fl &=&2^{(4-d)/2}\frac{\Gamma(d/2)}{(u\yw/v+1)(u\yw+1)}\int_0^1 dz\,dz'\, (zz')^{(d-4)/2}
\nn
\fl & & \times
(v-z-z'+1)^{-d/2}
\left(\frac{u\yw z}{v}+1\right)\left(\frac{u\yw z'}{v}+1\right)
\eea
and the replacement $u\to uv$ gives a similar expression for
$\gamma_d(uv\yw)\G(v,u\yw)/(\Tc^2\sigma_d)$.
%
We now insert these back into\eq{W_definition} to obtain
\bea
\fl \frac{2\mu_d u W}{\Tc^2\sigma_d}&=&
\frac{2^{(4-d)/2}\Gamma(d/2)}{u\yw+1}
\left[
\int_1^{\infty} dv\, v^{-(d+4)/4}
\frac{1}{(u\yw/v+1)^{1/2}}\sc{M}\!\left(\textstyle\frac{xv}{\yw},\frac{u\yw}{v}\right)\right.
\nn
\fl & & \times \left.
\int_0^1 dz\,dz'\, (zz')^{(d-4)/2}(v-z-z'+1)^{-d/2}
\left(\frac{u\yw z}{v}+1\right)\left(\frac{u\yw z'}{v}+1\right)\right.
\nn
\fl
& &{}+{}\left.
\int_1^{x/\yw} dv\,v^{(4-d)/4}
\frac{1}{(u\yw v+1)^{1/2}}
\sc{M}\!\left(\textstyle\frac{x}{v\yw},uv\yw\right)\right.
\nn
\fl & &\times\left.
\int_0^1 dz\,dz'\, (zz')^{(d-4)/2}
(v-z-z'+1)^{-d/2}
\left(u\yw z+1\right)\left(u\yw z'+1\right)
\right]
\eea
So far our calculation of $C^{(2)}$ applies for generic $x$; to
make more progress we consider again the large-$x$ behaviour.
In the first $v$-integral one can use directly the asymptotic
form\eq{M_asymptotics} of $\sc{M}$;
for the second integral one can show as in the fully magnetized
case~\cite{AnnSol06} that the same replacement can be made and 
the upper integration limit sent to infinity thereafter. This gives 
\bea
\fl \frac{2\mu_d W}{\Tc^2\sigma_d}&=&
\frac{d-2}{2} 2^{(4-d)/2}\Gamma(d/2)
x^{(2-d)/4}\frac{\yw^{(d-4)/4}}{u^{3/2}(u\yw+1)}\left[\int_1^{\infty} dv\,
v^{-d/2}
\int_0^1 dz\,dz'\, (zz')^{(d-4)/2}\right.
\nn
\fl & & \times\left.
(v-z-z'+1)^{-d/2}
\left(\frac{u\yw z}{v}+1\right)\left(\frac{u\yw z'}{v}+1\right)\right.
\label{W_asymptote}
\\
\fl & & {}+{}\left.
\int_1^{\infty} dv\,
\int_0^1 dz\,dz'\, (zz')^{(d-4)/2}
(v-z-z'+1)^{-d/2}
\left(u\yw z+1\right)\left(u\yw z'+1\right)\right]
\nonumber
\eea
Then from\eq{cm2_sc_m} one has
\bea
\fl C^{(2)}(t,\tw)&=&
\left(\frac{d-2}{2}\right)^2 \frac{2^{(4-d)/2}\Gamma(d/2)}{u^{1/2}(u+1)^{3/2}}\frac{\Tc^2\sigma_d \tw}{2\mu_d}x^{-(d+2)/4}
\int_0^1 d\yw\, \yw^{(d-2)/2}(1-\yw)^{(d-4)/2}
\nn
\fl & &\times
\left[\int_1^{\infty} dv\,\ldots + 
\int_1^{\infty} dv\, \ldots
\right]
\eea
where the $v$-integrals are as in\eq{W_asymptote}.
Carrying out the $\yw$-integral, this can be written as
\bea
\fl C^{(2)}(t,\tw)&=&
%
\left(\frac{d-2}{2}\right)^2
\frac{\Gamma((d+4)/2)}
{\Gamma((4-d)/2)\Gamma(d+1)}\,\frac{u^{3/2}\tw x^{-(d+2)/4}}{(u+1)^{3/2}}\left[
V_d+\frac{1}{u}V_d'+\frac{1}{u^2}V_d''\right]
\label{asymptotic_c_m2}
\eea
if we define $V_d$ as in the fully magnetized case, see~\ref{ref:corrections}, 
\be
\fl V_d=\int_1^{\infty}dv\,(v^{-(d+4)/2}+1)\int_0^1dz\,dz'\, (zz')^{(d-2)/2}(v-z-z'+1)^{-d/2}
\label{Vd}
\ee
and introduce also the analogous quantities
\be
\fl V_d'=\frac{4d}{d+2}\int_1^{\infty}dv\,(v^{-(d+2)/2}+1)\int_0^1dz\,dz'\, z^{(d-4)/2}z'^{(d-2)/2}(v-z-z'+1)^{-d/2}
\label{Vdp}
\ee
\be
\fl V_d'' =\frac{4(d-1)}{d+2} \int_1^{\infty}dv\,(v^{-d/2}+1)\int_0^1 dz\,dz'\, (zz')^{(d-4)/2}(v-z-z'+1)^{-d/2}
\label{Vds}
\ee
We have also used in\eq{asymptotic_c_m2} the explicit
expression~\cite{AnnSol06}
\be
\fl\frac{\Tc^2\sigma_d}{2\mu_d}=\frac{-2\Tc}{(d-2)(4-d)2^{(2-d)/2}\Gamma((4-d)/2)\Gamma((d-4)/2)\Gamma((d-2)/2)}
\ee

With the results\eq{asymptotic_R},\eq{asymptotic_c_m1} 
and\eq{asymptotic_c_m2} for the magnetization response and correlation
in the limit of long, well-separated ($x\gg 1$) times, we can finally
compute the asymptotic FDR as
\bea
\fl X^{\infty}&=&
\left(\frac{d-2}{2}u+1\right)(u+1)^3
\left\{2\left[\left(\frac{d(u+1)}{4}-\frac{3u}{2}\right)P_3(u)+u(u+1)P_3'(u)\right]
\right.
\nn
\fl & &{}+{}\left(\frac{d-2}{2}\right)^2\frac{\Gamma((d+4)/2)}{\Gamma((4-d)/2)\Gamma(d+1)}
\nn\fl & &\times\left.
\left[uP_2(u)\left(\frac{(d+12)(u+1)}{4}
-\frac{3u}{2}\right)-u(u+1)P_1(u)\right]\right\}^{-1}
\label{Xu_LargeX}
\eea
%
%
where we have defined the following $3$rd, $2$nd and $1$st order polynomials in $u$:
\bea
P_3(u)=Au^3+Bu^2+Cu+D\nn
P_2(u)=V_d u^2 +V_d' u+V_d''\nn
P_1(u)=V_d' u+ 2V_d''
\eea
The general structure of the asymptotic FDR is thus as for $d>4$, \ie\
a ratio of $4$th order polynomials in $u$. One can easily check 
that as $d\to 4$ the coefficients continuously approach those for
$d>4$, as they should. Also, the $u\ll 1$-limit of\eq{Xu_LargeX} retrieves the 
prediction for the unmagnetized case in $d<2<4$~\cite{GodLuc00b,AnnSol06}
\be
X^{\infty}=\frac{2}{dD}=\frac{d-2}{d}
\ee
Conversely, for $u\gg 1$ one has
\be
X^{\infty}=\frac{d-2}{2}\left[\frac{d+6}{2}A+\left(\frac{d-2}{2}\right)^2\frac{\Gamma((d+4)/2)}{\Gamma((4-d)/2)\Gamma(d+1)}\frac{d+6}{4}V_d\right]^{-1}
\ee
which stated in this form agrees with our earlier result for the
fully magnetized case~\cite{AnnSol06}. (The error was in an incorrect
expression for $V_d$; see~\ref{ref:corrections}.)
In this regime the asymptotic FDR interpolates between $X^\infty=1/2$
for $d=2$ (as can be shown by using that $V_d\sim 2/(d-2)$ to leading
order~\cite{AnnSol06}) and $X^{\infty}=4/5$ for $d=4$. As is required
by continuity with the situation for $d>4$, the contribution 
from $C^{(2)}$ vanishes as $d\to 4$, for any $u$. 
An $\epsilon=4-d$-expansion of\eq{Xu_LargeX} yields
\bea
X^{\infty}(u)&=&\frac{4(u+1)^4}{8+5u+30u^2+20u^3+5u^4}\nn
&&{}-{}2(u+1)^3
\frac{144+216u+48u^2+160u^3+95u^4+19u^5}{9(8+20u+30u^2+20u^3+5u^4)^2}
\,\epsilon
\label{Xinf_exp}
\eea
which in the unmagnetized ($u\ll 1$) and fully magnetized ($u\gg 1$)
limits reduces to $X^{\infty}(u=0)=1/2-\epsilon/8$ and
$X^{\infty}(u\to\infty)=4/5-(19/450)\epsilon$, respectively. The
former value agrees with the well-known result $X^\infty=(d-2)/d$ for
coarsening in the spherical model~\cite{GodLuc00b,AnnSol06} or the
$O(n\to\infty)$ model~\cite{CalGam02c} from an unmagnetized
state. The latter, corrected, value now also
agrees with the RG calculations for the longitudinal
fluctuations of the $O(n\to\infty)$ model~\cite{CalGam07}.

One interesting and unexpected feature of\eq{Xu_LargeX} and its
expansion\eq{Xinf_exp} is that the approach to the large-$u$ limit is
non-monotonic: for $d$ close to 4, $X^{\infty}(u)$ slightly overshoots
the limit value ``plateau'' and then decays down to it, signalling the
presence of a weak maximum. Expanding\eq{Xinf_exp} for large $u$ and
subtracting off its $u\to\infty$ asymptote, one sees that the
deviation from the plateau is controlled, to leading order, by two
terms with opposite signs, scaling respectively as $\epsilon/u^2$ and
$-1/u^4$. The maximum occurs where these two terms
compete, that is for $u\sim 1/\sqrt{\epsilon}$, or
$\bar{u}=u\sqrt{\epsilon}=\order{(1)}$.  Its height above the plateau
then scales as $\epsilon^2$. To get the scaling function determining
the shape of the maximum, we therefore normalize the deviation of $X^{\infty}$
from the large-$u$ plateau by $\epsilon^2$ and define \be D(\bar{u})\equiv
\lim_{\epsilon\rightarrow
  0}\frac{X^{\infty}(u=\bar{u}/\sqrt{\epsilon})-\lim_{u\to\infty}X^{\infty}(u)}{\epsilon^2}
=\frac{2(-18+5\bar{u}^2)}{75 \bar{u}^4}
\label{deviation}
\ee
This scaling function has its maximum at the finite value
$\bar{u}=6/\sqrt{5}$, as expected, and is positive for
$\bar{u}>\sqrt{18/5}$.

\begin{figure}
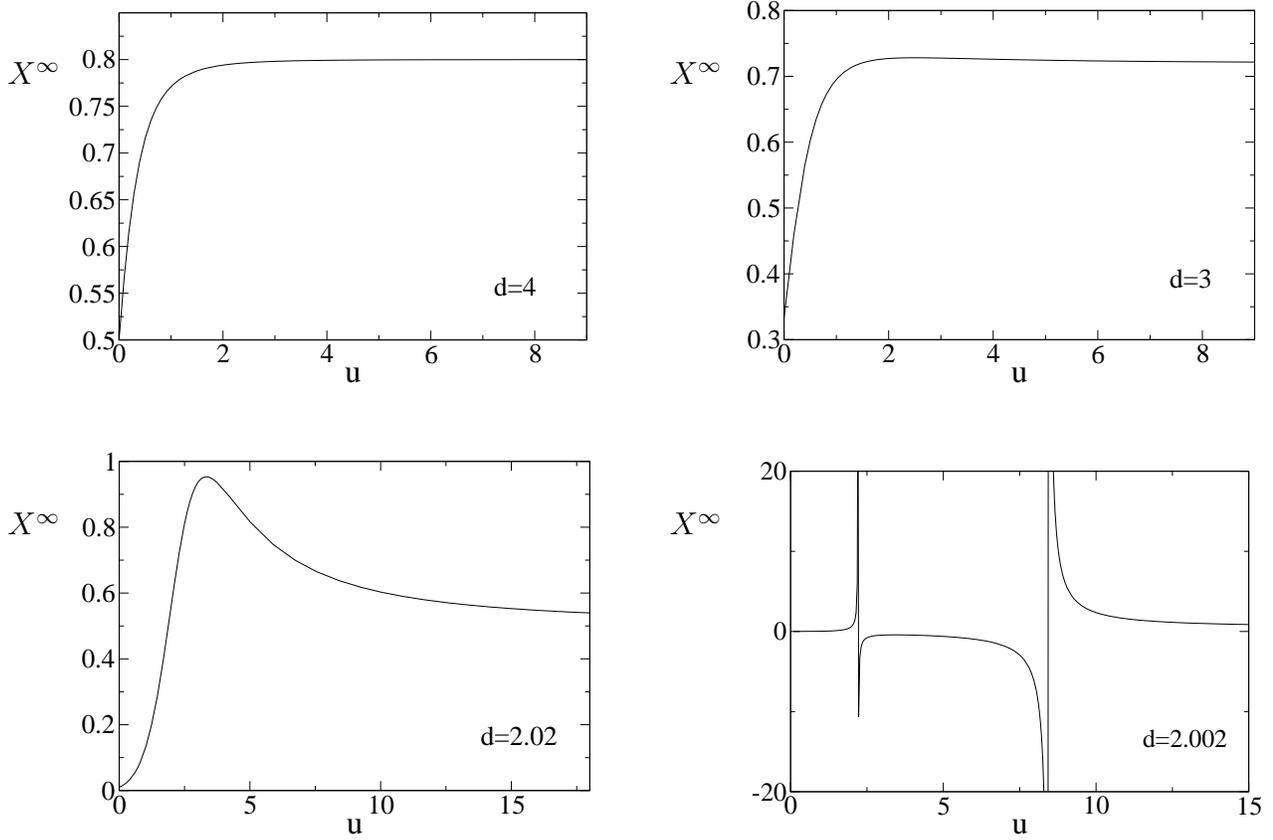

\setlength{\unitlength}{0.40mm}
\begin{picture}(230,300)(-100,0)
\put(-95,150){\includegraphics[width=173\unitlength]{Xinf_d4_vs_u.eps}}
\put(-115,250){$X^{\infty}$}
\put(130,150){\includegraphics[width=170\unitlength]{Xinf_d3_vs_u.eps}}
\put(105,250){$X^{\infty}$}
\put(-91,0){\includegraphics[width=170\unitlength]{Xinf_d202_vs_u.eps}}
\put(-115,100){$X^{\infty}$}
\put(132,0){\includegraphics[width=170\unitlength]{Xinf_d2002_vs_u.eps}}
\put(105,100){$X^{\infty}$}
\end{picture}
\caption{Asymptotic FDR $X^{\infty}$ for the magnetization vs $u=\tw/\taum$, 
for a few dimensions between $2$ and $4$ as indicated. As $d$
decreases, $X^\infty(u)$ develops an increasingly pronounced maximum
which eventually (see bottom right graph for $d=2.002$) 
turns into two
poles separated by a region of negative $X^\infty$.}
\label{fig:Xinf_dbelow4_vs_u}
\end{figure}

Looking next at dimensions further away from $d=4$,
Figure~\ref{fig:Xinf_dbelow4_vs_u} shows  
numerical values of $X^{\infty}$ for finite $u$ for a few 
dimensions $d$ between $2$ and $4$. $X^{\infty}$
converges to the fully magnetized value (which is near
$1/2$ for $d\approx 2$) for large $u$ and to the unmagnetized asymptotic FDR
$(d-2)/d$ for $u\rightarrow 0$ as it should. As anticipated, the interpolation
between these two limits is not, as in $d>4$, monotonic: 
$X^\infty$
initially increases with $u$ but ``overshoots'' its asymptotic
limit. This phenomenon
becomes more and more pronounced as $d\to 2$. For $d$ very close to 2,
finally, the maximum turns into two poles in $X^\infty(u)$, with
$X^\infty$ being negative in between.

The first pole is relatively straightforward to analyse from\eq{Xu_LargeX}.
One needs the dependence on $\delta=(d-2)/2$
of (\ref{A}--\ref{D}) and (\ref{Vd}--\ref{Vds}) for $\delta\rightarrow 0$. 
To leading order one finds $A=3\delta^2/4$, $B=3\delta/2$, 
$C=1$, $D=1/\delta$. The small $\delta$-limits of $V_d$, $V_d'$ and 
$V_d''$ one gets from (\ref{Vd}--\ref{Vds})
by noticing that the $v$-integrals become dominated by their
large $v$ tails as $\delta\to 0$, giving $V_d=1/\delta$,
$V_d'=2/\delta^2$ and $V_d''=1/\delta^3$.
Gathering these results, equation\eq{Xu_LargeX} becomes to leading order
\be
X^{\infty}=(u+1)^3\left[
(-2u+1)D+\frac{3}{2}\delta^2 u V_d''
\right]^{-1}
=\delta\frac{2(u+1)^3}{2-u}
\label{delta0_fixed_u}
\ee
which approaches for $u\rightarrow 0$ 
the unmagnetized limit $X=\delta+\order(\delta^2)$ as it should. For larger $u$ we read
off that there is a pole at $u=2$ beyond which $X^\infty$ is
negative. In fact, the expression\eq{delta0_fixed_u} shows that
$X^\infty\approx -2\delta u^2$ for large $u$ whereas we expect
convergence to the known limit $X^\infty=1/2$. The reason is that the 
limits $u\rightarrow\infty$ and $\delta\rightarrow 0$ 
do not commute: the approach to the eventual asymptotic value takes
place on a scale of values of $u$ that diverges as $\delta\to0$.

\begin{figure}
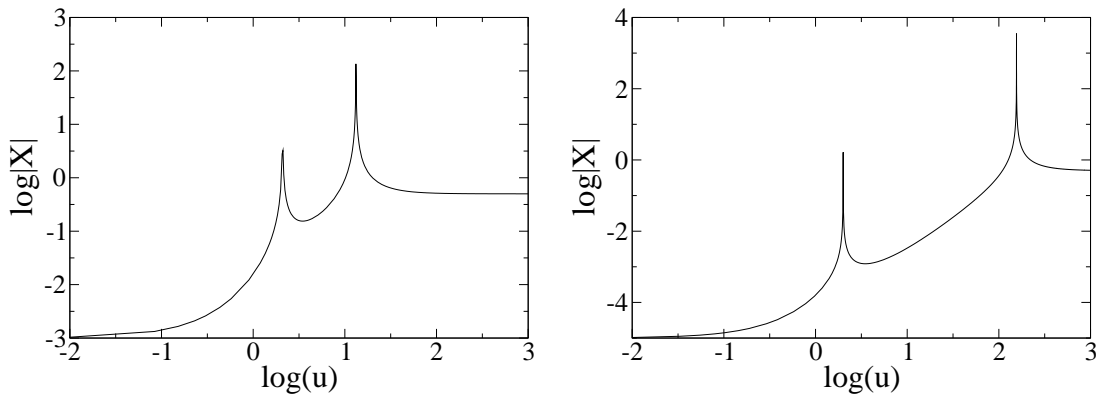

\centerline{\includegraphics[width=7.0cm,clip=true]{LogLog_Xinf_vs_u_de-3.eps}
\hspace{0.2cm}
\includegraphics[width=7.0cm,clip=true]{LogLog_Xinf_vs_u_de-5.eps}}
\caption{Log-log plot of the absolute value of the asymptotic FDR 
$X^{\infty}$ vs $u$, 
for $\delta=10^{-3}$ (left) and $\delta=10^{-5}$ (right). 
Note that the second pole moves to the right as $\delta$ decreases,
with the expected proportionality $u\sim \delta^{-1/2}$. Here and in
the following plots, $\log\equiv\log_{10}$.
\label{fig:LogLog_Xinf_dbelow4_vs_u}
}
\end{figure}

The form of the response 
function as given in\eq{R_generic_x} would suggest that the appropriate
diverging $u$-scale to consider is $u\sim 1/\delta$: in this regime
the two terms in the square brackets in\eq{R_generic_x}, which cancel
exactly for $u\to\infty$ and $d\to 2$, still give a leading order
cancellation. However, one finds with a bit of algebra that the limit
as $\delta\to 0$ of $X^\infty$, taken at fixed $u'=u\delta$, is simply the
constant asymptotic value $X^\infty=1/2$. The crossover to this
asymptotic regime must therefore take place on shorter timescales $u$.
To explore this, we need to look more closely at
the polynomial structure of $X^{\infty}$.
As observed, $X^{\infty}$ 
can be written as the ratio of fourth order polynomials,
\be
X^{\infty}=
\frac{a u^4+b u^3 + c u^2 + d u +e}{a' u^4+b' u^3 + c' u^2 + d' u +e'}
\label{polynomial_ratio}
\ee
The coefficients can be computed in the limit 
$\delta\rightarrow 0$ and their leading terms evaluate to
\bea
\begin{array}{ll}
a=\delta \hspace{1cm} & a'=2\delta\\
b=1+3\delta \hspace{1cm} & b'=2+\frac{9}{2}\delta\\
c=3+3\delta \hspace{1cm} & c'=\frac{11}{2}+\frac{5}{4}\delta\\
d=3+\delta \hspace{1cm} & d'=-\frac{1}{2\delta}+\frac{9}{4}\\
e=1 \hspace{1cm} & e'=1+\frac{1}{\delta}\\
\end{array}
\eea
We now consider values of $u$ diverging as some generic power of
$\delta$, $u=u''\delta^{-\beta}$. In the limit $\delta\to 0$ a number
of terms can then be dropped: \eg\ $cu^2$ in the numerator is always
subleading compared to $bu^3$ because both $b$ and $c$ are order unity
but $u\gg 1$ for $\delta\ll 1$. For the same reason the terms
proportional to $d$, $e$, $c'$ and $e'$ can never be
leading. Retaining only the other, potentially leading, terms gives
\be
X^{\infty}=\frac{\delta^{1-4\beta}
u''^4+\delta^{-3\beta}u''^3}{2\delta^{1-4\beta}
u''^4+2\delta^{-3\beta}u''^3-\frac{1}{2}\delta^{-1-\beta}u''}
\label{Xinfty_general_small_delta}
\ee
Comparing powers of $\delta$ shows that the only values of $\beta$
for which in 
the limit $\delta\to 0$ more than one term
survives in either numerator or denominator are $\beta=1/2$, $2/3$
and $1$. The competing terms at $\beta=2/3$ are both subleading
so this case is uninteresting. Only $\beta=1/2$ therefore remains as
a non-trivial exponent value to analyse.
One then has explicitly
$u''=u\delta^{1/2}$ and the surviving terms
in\eq{Xinfty_general_small_delta} are
\be
X^{\infty}=\frac{\delta^{-3/2}u''^3}
{2\delta^{-3/2}u''^3-\frac{1}{2}\delta^{-3/2}u''}
=\frac{2u''^2}{4u''^2-1}
\ee
This result matches the magnetized limit $X_m^{\infty}=1/2$ for large
$u''$ as it should; for $u''=1/2$ it has a pole and for small $u''$ it
is negative and small. In the latter regime,
$X^\infty=-2u''^2=-2\delta u^2$ also matches smoothly with the large-$u$
limit of\eq{delta0_fixed_u} as it
should. Figure~\ref{fig:LogLog_Xinf_dbelow4_vs_u} demonstrates this
behaviour by showing on a logarithmic scale the absolute value of
$X^{\infty}$ versus
$u$. As $\delta$ decreases, the second pole moves to larger
$u=1/(2\delta^{1/2})$ as expected while the first one occurs at a finite
limiting value of $u$, $u=2$.
%

It would clearly be desirable to understand in more detail the origins of the 
highly non-trivial  
behaviour of the asymptotic magnetization FDR for $d$ near $2$, and to
ascertain how this behaviour is reflected in the corresponding FD plots.
The response\eq{R_generic_x} is always positive, so from the
definition of the FDR in\eq{FDR_above} singularities in $X$ can arise
only from zeros in $\partial_{\tw} C$, \ie\ from a non-monotonic dependence of the
magnetization correlator on $\tw$. As we will show, this
non-monotonicity arises because the equal-time correlator has a
pronounced maximum around $u=\tw/\taum=1$, and this large variance of
the magnetization fluctuations leaves its imprint in the two-time
correlator as a weak maximum.

The main difficulty we now face is to obtain the behaviour of the
correlator also for finite $x$ rather than just $x\gg 1$. This is made
possible by the following observation: in the $\delta=(d-2)/2\to 0$-limit,
the scaling function $\sc{M}(x/y,uy)$ from\eq{scM_xyuy} develops a
non-integrable singularity at $y=x$. This concentrates the weight of
any integrand into this region, so that 
for any function $f(y,u)$ which is smooth at $y=x$ 
\be
\int_0^x dy\, \sc{M}\left(\frac{x}{y},uy\right)f(y,u)\to \frac{ux^{1/2}}{ux+1}f(x,u)
\label{delta_approximation}
\ee
as $\delta\to 0$, \ie\ $\sc{M}(x/y,uy)$ acts effectively as 
$\sc{M}(x/y,uy)=[ux^{1/2}/(ux+1)]\delta(x-y)$.
The same observation applies to $\sc{M}(1/\yw,u\yw)$, which is obtained
by setting $x=1$ and $y=\yw$.
These approximations will yield the leading terms in the correlator
in the limit $\delta\rightarrow 0$. We will also use them
for dimensions slightly above $2$ to explore 
numerically the $x$-dependence of the correlator and the resulting FD
behaviour. Even though the results here no longer have the character
of a systematic expansion in $\delta$, they 
will give insights into the non-trivial $d$-dependence of the 
FD plots for $d$ close to $2$.
 
Using the approximation\eq{delta_approximation}, 
the contributions to $C^{(1)}$ coming from $S$ vanish because the
first argument of $\sc{M}(x/y,uy)$ is always $\geq x>1$. 
The remaining term $F$ is given explicitly in\eq{F1} and taking the
$\delta\to 0$ limit gives
\be
\frac{C^{(1)}(t,\tw)}{\Tc t}=\frac{2x^{-1/2}}{\delta(ux+1)^{3/2}(u+1)^{3/2}}\left(1+u\delta+u^2 \delta^2\right)
\label{c_m1_d2}
\ee
We note that a naive application of the $\delta$-approximation
explained above would in this case give an incorrect result, because
it produces a leading order cancellation of terms when $u\sim
1/\delta$. The remaining subleading term is then of the same order as the first
correction to the $\delta$-approximation. One can nevertheless check
that the second contribution, $S$, to $C^{(1)}$ always remains negligible compared to $F$ because it is subject to a similar cancellation.

\begin{figure}
\setlength{\unitlength}{0.40mm} 
\begin{picture}(200,185)(-100,20)
\put(-90,30){\includegraphics[width=180\unitlength]{EqCm_delta0001_vs_utt.eps}}
\put(120,20){\includegraphics[width=160\unitlength]{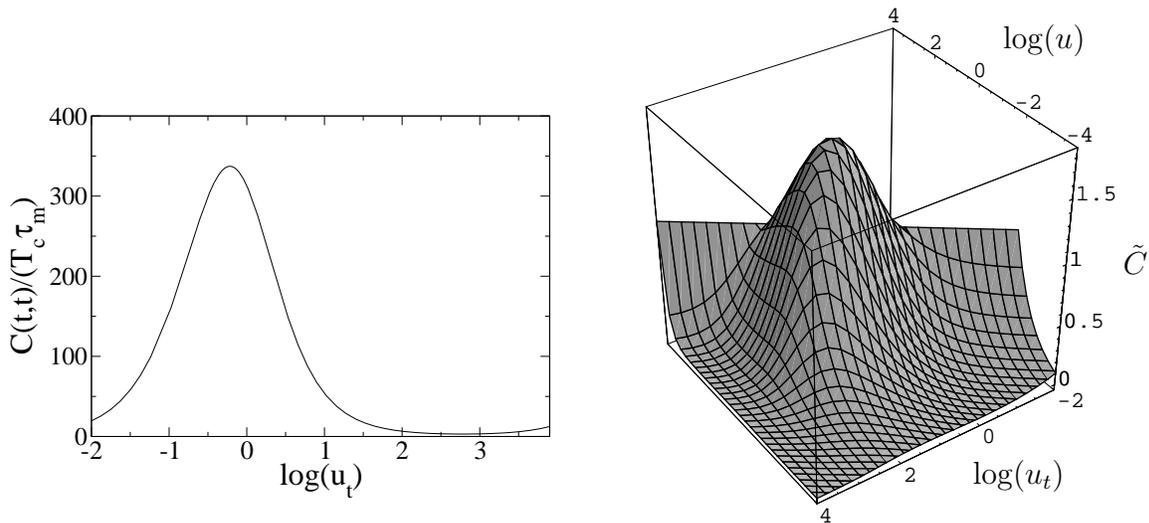}}
\put(240,180){$\log(u)$}
\put(280,105){$\tilde{C}$}
\put(230,35){$\log(u_t)$}
\end{picture}
\caption{Left: The equal-time correlator $C(t,t)/(\Tc\taum)$
versus $\log(u_t)$ for 
$\delta=10^{-3}$ shows pronounced non-monotonic behaviour.
Right: Normalized correlator $\tilde C$ for the same $\delta=10^{-3}$, plotted
versus $\log(u)$ and $\log(u_t)$. Note the non-monotonicities in the
$u$-dependence around $\log(u)=0$. 
}
\label{fig:correlator_d2}
\end{figure}

Next we want to compute $C^{(2)}$. 
In\eq{W_definition} the first integral again vanishes because the
first argument of $\sc{M}$ is always $>1$, so
\bea
\fl W&=&\frac{\Tc^2\sigma_d}{\mu_d}\left(\frac{x}{\yw}\right)^{(4-d)/4}\frac{x^{3-d/2}}{(u\yw+1)(ux+1)^{3/2}}\int_0^1 dz\,dz'\, (zz')^{(d-4)/2}
\nn
\fl & &\times\left(\frac{x}{\yw}-z-z'+1\right)^{-1}
\left(u\yw z +1\right)\left(u\yw z'+1\right)
\eea
which for small $\delta$ evaluates to
\bea
\fl W&=&\frac{\Tc^2\sigma_d}{\mu_d}\left(\frac{x}{\yw}\right)^{(4-d)/4}\frac{x^{3-d/2}}{(u\yw+1)(ux+1)^{3/2}}
\left\{
u^2\yw^2\left[
\frac{x}{\yw}\ln\left(1-\frac{\yw^2}{x^2}\right)\right.\right.
\nn
\fl & & {}+{}\left.\left.\ln\left(\frac{x+\yw}{x-\yw}\right)
\right]+\frac{2u\yw}{\delta}\ln\left(\frac{x+\yw}{x}\right)
+\frac{1}{\delta^2}\frac{\yw}{x+\yw}
\right\}
\eea
The $\yw$-integral in\eq{cm2_sc_m} can be performed by again using the 
$\delta$-approximation and one finally gets 
\bea
\frac{C^{(2)}(t,\tw)}{\Tc t}&=&\frac{u
x^{1/2}}{\delta(u+1)^{3/2}(u x+1)^{3/2}}
\left\{\frac{1}{x+1}+2u\delta\ln\left(\frac{x+1}{x}\right)
\right.\nn
&&{}+{}\left.
u^2\delta^2\left[x\ln\left(1-\frac{1}{x^2}\right)+\ln\left(\frac{x+1}{x-1}\right)\right]
\right\}
\label{c_m2_d2}
\eea
Adding the results\eq{c_m1_d2} and\eq{c_m2_d2} gives the magnetization
correlation function for generic $x$ and $d$ close to $2$.  By pulling
a factor of $x$ into the curly bracket of the latter, one sees that
the resulting terms in the bracket are of $\order(1)$, $\order(\delta u)$ and
$\order(\delta^2 u^2)$ for all $x$, as in\eq{c_m1_d2}. But the
prefactor in\eq{c_m2_d2} is larger by a factor of $u$, so for $u\gg 1$
we can neglect $C^{(1)}$ against $C^{(2)}$. This implies that
we can always drop the $\order(\delta u)$ and $\order(\delta^2 u^2)$
terms in $C^{(1)}$: either $u=\order(1)$, and then they are subleading
compared to the $\order(1)$ term in $C^{(1)}$, or $u\gg 1$ and they
are small compared to the corresponding terms in $C^{(2)}$.
Changing then also to a normalization with $1/\taum$ instead of $1/t$, we can
express the leading order terms for $\delta\to0$ of the magnetization
correlator as a function of the scaled times $u=\tw/\taum$ and
$u_t=t/\taum=xu$ in the form 
\bea
\frac{C(t,\tw)}{\Tc \taum}&=&\frac{u^{1/2}u_t^{3/2}
}{\delta(u+1)^{3/2}(u_t+1)^{3/2}}
\left\{\frac{2}{u_t}+\frac{u}{u+u_t}+2u\delta\ln\left(\frac{u+u_t}{u_t}\right)
\right.\nn
&&{}+{}\left.
u^2\delta^2\left[\frac{u_t}{u}\ln\left(1-\frac{u^2}{u_t^2}\right)
+\ln\left(\frac{u+u_t}{u_t-u}\right)\right]
\right\}
\eea
The equal-time correlator is then
obtained by taking the limit $u\rightarrow u_t$: 
\be
\frac{C(t,t)}{\Tc \taum}=\frac{u_t(4+u_t
+4\ln2\, \delta u_t^2
+4\ln2\, \delta^2 u_t^3)}{2\delta (u_t+1)^3}
\label{Cm_dnear2}
\ee
Evaluating this numerically for small $\delta$ as shown in
Fig.~\ref{fig:correlator_d2}, we see that it is
non-monotonic in $u_t$ as anticipated.
In fact, the expression\eq{Cm_dnear2} shows directly that the height
of the peak at $u_t=\order(1)$ diverges as $1/\delta$ for $\delta\to
0$, whereas for $u_t=\order(1/\delta)$ the result is of order unity.
%
%
On the right of Fig.~\ref{fig:correlator_d2} we demonstrate that the
peak in the equal-time correlation function does indeed cause
corresponding non-monotonic behaviour in the (normalized) two-time
correlator $\tilde{C}(t,\tw)=C(t,\tw)/C(t,t)$ in the region where
$u=\order(1)$. Notice that in $\tilde C$ the prefactors $\Tc \taum$ that
we have isolated on the left of the expressions above cancel and we
obtain a function of only $u_t$ and $u$: 
\bea
\fl \tilde{C}(u_t,u)&=&\left(\frac{u_t+1}{u+1}\right)^{3/2}
\frac{2 u^{1/2} u_t^{1/2}}{4 + u_t + 4\ln2\,\delta u_t^2 + 4\ln2\,\delta^2 u_t^3}
\Biggl\{\frac{2}{u_t}+\frac{u}{u+u_t}+2u\delta\ln\left(\frac{u+u_t}{u_t}\right)
\nn
\fl
&&{}+{}
u^2\delta^2\left[\frac{u_t}{u}\ln\left(1-\frac{u^2}{u_t^2}\right)
+\ln\left(\frac{u+u_t}{u_t-u}\right)\right]
\Biggr\}
\label{Ctildem_dnear2}
\eea
\begin{figure}
\setlength{\unitlength}{0.40mm}
\begin{picture}(200,145)(-100,0)
\put(10,10){\includegraphics[width=180\unitlength,clip]{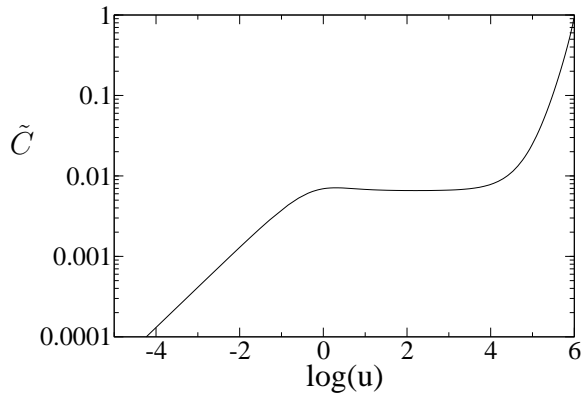}}
\put(0,90){$\tilde{C}$}
\end{picture}
\caption{Normalized correlator versus $\log(u)$ at fixed $u_t=10^6$ and 
$\delta=10^{-5}$.
}
\label{fig:correlator_non_mono}
\end{figure}
%

We now analyse more closely the nature and scaling of the
non-monotonicities of the normalized two-time correlator. For $u_t$ of
order unity, all terms involving powers of $\delta$ can be dropped
in\eq{Ctildem_dnear2}. The resulting function is monotonically
increasing in $u=\tw/\taum=0\ldots u_t$ for $u_t>\sqrt{13}-3\approx0.61$. For
larger $u_t$ it has a maximum in $u$ whose position shifts from
$\sqrt{13}-3$ to an asymptotic limit of 2 as $u_t$ increases. If one keeps
the terms that are 
subleading in $\delta$, one sees that $\tilde{C}$ contains a
contribution scaling as $\delta^2(u_t-u)\ln(u_t-u)$. This yields
a term $-\delta^2\ln(u_t-u)$ in the $u$-derivative of $\tilde{C}$
which diverges to $+\infty$ as $u\to u_t$, and so gives a positive
sign for the derivative in 
the limit. Whenever $\tilde{C}$ has a maximum as a function of $u$ it
therefore also has an associated minimum, but this is essentially
undetectable as it occurs extremely close to $u_t$, for $u_t-u\sim
\exp(-{\rm const}/\delta^2)$.

\begin{figure}
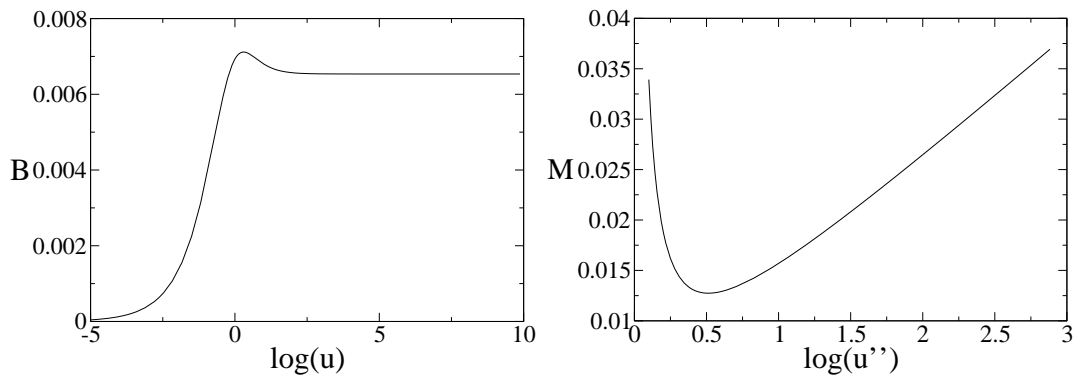

\includegraphics[width=7.0cm,clip=true]{bump_utdel10_vs_ulog.eps}
\includegraphics[width=7.0cm,clip=true]{minimum_utdel10_vs_usqrt.eps}
\caption{Position and scaling of the maximum and the minimum of the 
normalized two-time correlator as a function of $u$ or
$u''=u\delta^{1/2}$ for a 
fixed value of $u_t\sim1/\delta$; for the example in the plot we have
taken $u_t\delta=u_t'=10$.} 
\label{fig:bump_minimum}
\end{figure}

Moving to larger $u_t$ of order $1/\delta$, the position of the
minimum in $\tilde{C}$ becomes clearly separate from $u_t$.
An example of this is shown in Fig.~\ref{fig:correlator_non_mono},
which graphs the normalized correlator as a function 
of $\log(u)$ for a fixed value of $u_t$ with $u_t\delta=10$: 
one discerns a small maximum followed by a broad minimum. (Numerically, one
finds that these features merge once $d$ gets sufficiently far above
$2$, restoring monotonicity.) The maximum in $u$ is, for $u_t\sim
1/\delta$, always located at $u=2$; this matches the behaviour
discussed above for
large $u_t$ of $\order(1)$. Explicitly, if we let $\delta$ tend to $0$ 
in the normalized correlator at fixed $u$ and $u_t'=u_t\delta$ we get
\be
B(u_t',u)\equiv \lim_{\delta\rightarrow 0}\tilde C(u_t'/\delta,u)
=\frac{2\sqrt{u}(u+2)}{(u+1)^{3/2}(1+4\ln2\, u_t' +4\ln2\,u_t'^2)}
\label{bump}
\ee
The result is shown in Figure~\ref{fig:bump_minimum} (left) and does
have a maximum at $u=2$ as anticipated. This value makes sense since
it was also the point where the asymptotic FDR $X^\infty(u)$ diverges
for $d$ close to 2. (Given that $u_t\sim 1/\delta$ we are
automatically in the asymptotic regime $u_t\gg u$.)

The position $u_{\rm min}$ of the corresponding minimum of $\tilde{C}$
as a function of $u$ is somewhat more subtle. For $u_t'<1/2$, it is
located at $u_{\rm min}\sim 1/\delta$, \ie\ $u_{\rm min}<u_t$ but with
the two values being of the same order. As $u_t'\to 1/2$ from below, $u_{\rm
min}\delta \to 0$; for even larger values of $u_t'$, one finds a
different scaling $u_{\rm min}\sim \delta^{-1/2}$ so that always
$u_{\rm min}\ll u_t$.
To find the minimum position in this regime we need to fix
$u''=u{\delta}^{1/2}$ in the 
normalized correlator when letting $\delta\to 0$. The typical values
of $\tilde C$ in this regime
turn out to be only $\order(\delta^{1/2})$ above the plateau
$B(u_t',u\to\infty)$ so we subtract off the latter and divide by
$\delta^{1/2}$ to define
\bea
M(u_t',u'')
&=&\lim_{\delta\rightarrow 0}\frac{\tilde{C}(u_t'/\delta, u''/\delta^{1/2})
- B(u_t',u\to\infty)}
{\delta^{1/2}}
\\
&=&\frac{u_t'+u''^2(-2+4u_t')}{u''u_t'(1+4\ln2\, u_t'+4\ln2\, u_t'^2)}
\eea
We show a sample plot of this, for a specific value of $u_t'>1/2$, in
Fig.~\ref{fig:bump_minimum} (right).
The minimum of $M$ occurs at $u''_{\rm min}=[u_t'/(4u_t'-2)]^{1/2}$, which for 
large $u_t'$ yields $u''_{\rm min}=1/2$. This matches the
position of the second divergence of the asymptotic FDR $X^\infty(u)$,
as it should.

As explained above, the derivative of the two-time correlator is
always dominated by a logarithmically divergent term in the limit
$u\to u_t$; including prefactors, this reads $\partial_{\tw} C(t,\tw)
= -\Tc \delta\ln(u_t'-u')$ in the regime $u_t=u_t'/\delta$,
$u=u'/\delta$. The response function\eq{R_generic_x} is then dominated
by the same logarithmic terms:
\be
R(u_t',u')=\frac{\delta}{u_t'}\left[1-u_t'\ln\left(1-\frac{u'}{u_t'}\right)\right]
\approx -\delta\ln(u_t'-u')
\ee 
The last approximation, which gives the dominant term for $u'\to
u_t'$, shows that the initial (negative) slope of the FD plot is always
exactly equal to one. However, as $u_t'$ becomes small this becomes
undetectable because so does the logarithmic singularity in the correlator.

We are now in a position to analyse the magnetization FD plots for $d$
near 2. We consider normalized plots ($\tilde\chi$ vs $\tilde C$) as
in $d>4$, holding $u_t$ fixed for each plot as before to get a valid
connection with the FDR $X$ and varying $u$. We obtain $\tilde\chi$ by
numerical integration of\eq{R_generic_x}, according to the
definition\eq{cross_chi_dgt4}, and then dividing by\eq{Cm_dnear2}.
The asymptotic FDR $X^\infty(u)$ that we have calculated applies in
the limit $u_t\gg u$, corresponding to the region in the top left hand
corner of an FD plot. Starting from the top left corner ($u\to 0$) we
then expect to see in the FD plots the slopes varying as given by
$X^\infty(u)$: initially small (of $\order(\delta)$) and negative as
usual, then turning positive and of order unity, and finally negative
again. This S-shape should be present for large $u_t$; for
smaller $u_t$, only part of this variation will be accessible because 
$u\leq u_t$.

\begin{figure}
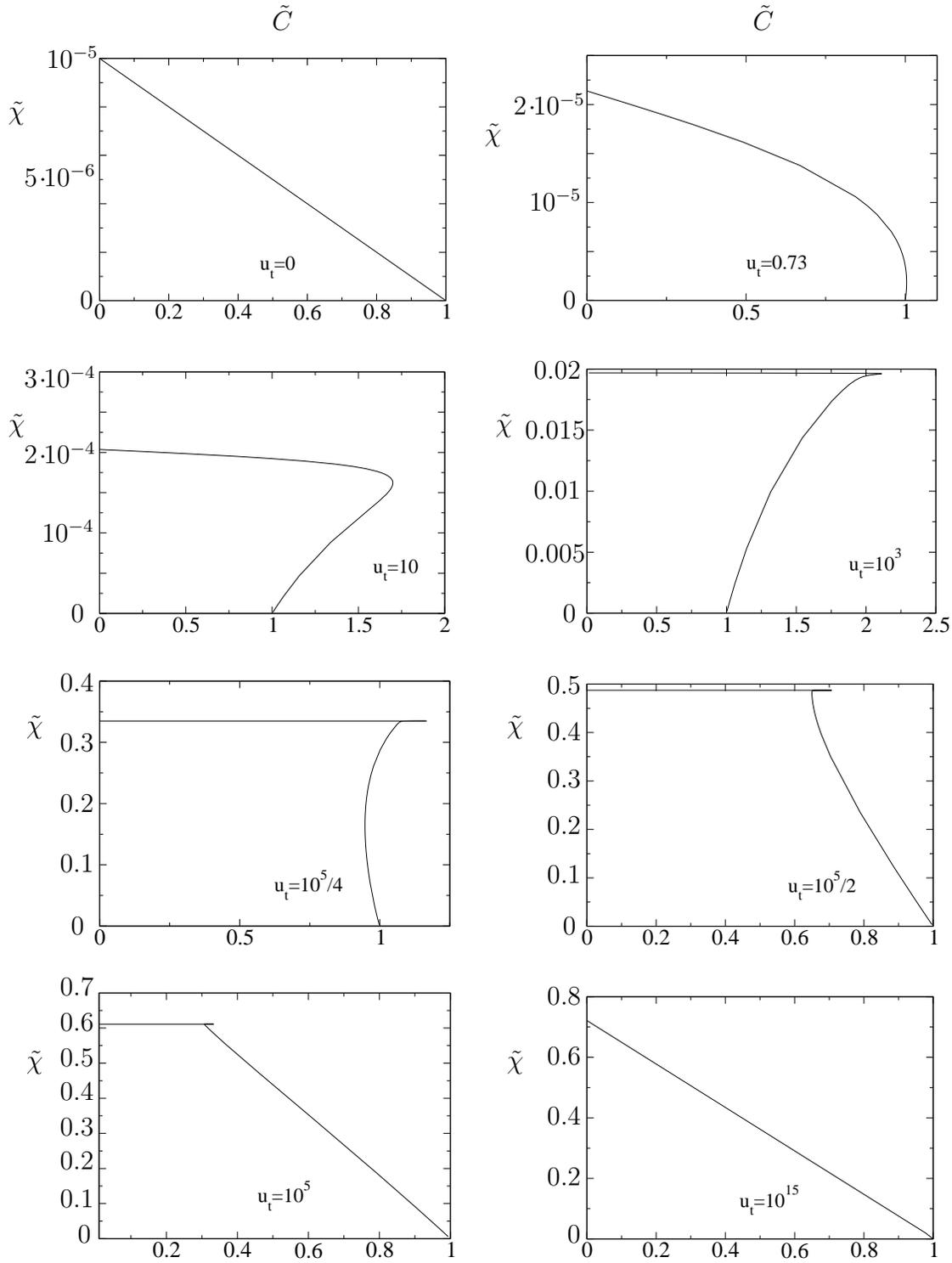

\setlength{\unitlength}{0.35mm}
\begin{picture}(300,580)(-100,20)
\put(-70,440){\includegraphics[width=160\unitlength]{FD_utt0_de-5.eps}}
\put(-91,553){$10^{-5}$}
\put(-97,500){$\!\!5\!\cdot \!\!10^{-6}$}
\put(-76,445){$0$}
\put(148,440){\includegraphics[width=160\unitlength]{FD_ut073_de-5.eps}}
\put(10,570){$\tilde{C}$}
\put(225,570){$\tilde{C}$}
\put(-108,530){$\tilde{\chi}$} 
\put(105,520){$\tilde{\chi}$} 
\put(120,532){$\!2\!\cdot \!\!10^{-5}$}
\put(127,489){$\!10^{-5}$}
\put(140,445){$0$}
\put(-70,300){\includegraphics[width=160\unitlength]{FD_utt1_de-5.eps}}
\put(-97,410){$\!\!3\!\cdot \!\!10^{-4}$} 
\put(-97,376){$\!\!2\!\cdot \!\!10^{-4}$} 
\put(-91,341){$\!10^{-4}$} 
\put(-80,305){$0$}
\put(148,300){\includegraphics[width=165\unitlength]{FD_utt3_de-5.eps}}
\put(129,414){$\!0.02$} 
\put(125,387){$\!0.015$} 
\put(129,360){$\!0.01$} 
\put(125,332){$\!0.005$} 
\put(140,305){$0$} 
\put(-108,390){$\tilde{\chi}$} 
\put(110,390){$\tilde{\chi}$}  
\put(-70,160){\includegraphics[width=160\unitlength]{FD_utt5_de-5quarti.eps}}
\put(-85,274){$0.4$} 
\put(-85,247){$0.3$} 
\put(-85,220){$0.2$} 
\put(-85,193){$0.1$} 
\put(-80,165){$0$} 
\put(148,160){\includegraphics[width=160\unitlength]{FD_utt5half_de-5.eps}}
\put(135,273){$\!0.5$} 
\put(135,252){$\!0.4$} 
\put(135,231){$\!0.3$} 
\put(135,210){$\!0.2$} 
\put(135,188){$\!0.1$} 
\put(140,165){$0$} 
\put(-100,255){$\tilde{\chi}$} 
\put(115,255){$\tilde{\chi}$}  
\put(-68,20){\includegraphics[width=160\unitlength]{FD_utt5_de-5.eps}}
\put(-85,138){$0.7$} 
\put(-85,122){$0.6$} 
\put(-85,106){$0.5$} 
\put(-85,91){$0.4$} 
\put(-85,75){$0.3$} 
\put(-85,60){$0.2$} 
\put(-85,43){$0.1$} 
\put(-77,27){$0$} 
\put(148,20){\includegraphics[width=160\unitlength]{FD_utt15_de-5.eps}}
\put(135,134){$\!0.8$} 
\put(135,107){$\!0.6$} 
\put(135,80){$\!0.4$} 
\put(135,53){$\!0.2$} 
\put(142,27){$0$} 
\put(-100,105){$\tilde{\chi}$} 
\put(115,105){$\tilde{\chi}$}  
\end{picture}
\caption{Normalized FD plots for $d$ close to $2$, 
showing normalized susceptibility 
$\tilde{\chi}$ versus normalized correlation $\tilde{C}$, for
$\delta=10^{-5}$ and increasing
values of $u_t$ as shown in each plot. 
Once the S-shape appears, it remains present for all
larger $u_t$ but gets squashed into a region in the top left hand
corner scaling as $1/(u_t\delta)^2$.
The slope where the plot meets the $y$-axis is always given by
$X^{\infty}(u=0)\approx\delta$.
}
\label{fig:FD_dclose2}
\end{figure}

For $d$ close to $2$ the above expectations are indeed borne out by
numerical evaluation as illustrated in Figure~\ref{fig:FD_dclose2} for
$\delta=10^{-5}$. As $u_t$ increases, we start from the fully linear
FD plot of the unmagnetized case, with negative slope $X=X^\infty(u=0)\approx
\delta$. A section of much larger $X$ then grows and eventually ``flips'' to
the right, producing a region of negative FDRs.  For much larger
values ($u_t\sim \delta^{-1}$)
the beginning of the FD plot (equal
times, where the plot meets the horizontal axis) eventually swings
back to the left to return to the conventional negative slope. The
initial slope of $-1$ also becomes visible. At this
stage the expected S-shape is complete; it then shrinks progressively towards
the top left corner as $u_t$ grows and the rest of the plot approaches
the close-to-linear shape~\cite{AnnSol06} for the fully magnetized
case. The region of the plot occupied by the ``S'' scales as
$1/u_t'^2=1/(u_t\delta)^2$ for $u_t\gg 1/\delta$. This is clear
from\eq{bump} which gives the typical values of $\tilde C$ at the
maximum and minimum, \ie\ at the right and left boundary of the ``S''.
The S-shaped region ends where the plot
meets the $y$-axis with an asymptotic slope that is $u_t$-independent:
this point corresponds to 
$u\to 0$, so we are always in the regime $u_t\gg u$ where the
asymptotic FDR $X^\infty$ applies and the negative slope is
$X^\infty(u=0)\approx\delta$. (In fact, this argument applies for any
$u_t$, whether or not an actual S-shape is present.)
%

The crossover between unmagnetized and fully magnetized behaviour can
also be seen from the $u_t$-dependence of the $y$-axis intercept of
the FD plot, which can be thought of as its ``axis ratio'' $Y$.
This is found from the large $x$ limit at fixed $u_t$ of the
susceptibility, multiplied by $\Tc$ and normalized by the equal-time correlator $C(t,t)$ from\eq{Cm_dnear2}.
The former is determined from the response function\eq{R_generic_x} by
integration, $\chi(t,\tw)=\int_{\tw}^t dt'\,R(t,t')$. Rescaling
$t'=zt$ gives
\bea
\fl \chi(t,\tw)&=&
\frac{t}{(u_t+1)^{1/2}}\int_{1/x}^1 dz\, z^{(d-4)/4}(u_t z+1)^{1/2}
\left[1-\frac{u_t}{u_t+1}(1-z)^{(d-2)/2}\right]
\eea
In the small $\delta$-limit at fixed $u_t=\order(1)$, the square
bracket simplifies to $1/(u_t+1)$ and the $z$-integral can be done
explicitly. Multiplying by $\Tc/C(t,t)$ gives for the axis ratio in
this regime
\be
Y = \delta\frac{2(u_t+1)^{3/2}}{4+u_t}
\left[\sqrt{1+u_t}+u_t^{-1/2}\ln(\sqrt{u_t}+\sqrt{1+u_t})\right]
\label{Y_small_ut}
\ee
This is of order $\delta$ as expected
from the FD plots in 
Fig.~\ref{fig:FD_dclose2}. For $u_t\to 0$ one gets $Y=\delta$ exactly,
consistent with the known results for the unmagnetized
case~\cite{AnnSol06}; for large $u_t$, on the other hand, $Y=2\delta
u_t$.

\begin{figure}
\centerline{\includegraphics[width=7.0cm,clip=true]{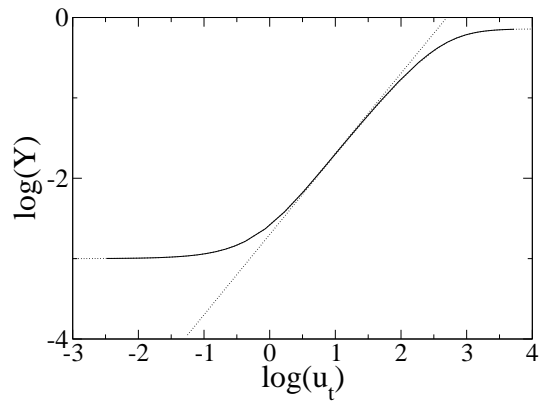}}
\caption{Axis ratio $Y$ of FD plot, given by the asymptotic normalized
susceptibility $\tilde\chi(u_t,u\to 0)$, versus $\log(u_t)$ 
for $\delta=10^{-3}$. The scaling functions\eq{Y_small_ut}
and\eq{Y_large_ut} are shown and match in the crossover regime ($1\ll
u_t \ll 1/\delta$) as expected. The dotted lines represent their
continuations towards larger and smaller $u_t$, respectively.
For $u_t=\order(1)$, $Y$ is $\order{(\delta)}$ as expected 
from the unmagnetized case, whereas for $u_t\gg 1/\delta$ we retrieve the 
magnetized limit $Y=1/(2\ln2)$.}
\label{intercept}
\end{figure}

In the regime $u_t\sim 1/\delta$ one finds similarly, by setting
$u_t=u_t'/\delta$ and taking $\delta\to 0$
\be
Y = \frac{2u_t'(1+u_t')}{1+4\ln 2\,u_t'+4\ln 2\,u_t'^2}
\label{Y_large_ut}
\ee
This is of order unity, again consistent with the FD plots shown
above. For $u_t'\ll 1$ it approaches $2u_t'$, matching the result from the
previous regime, while for $u_t'\to\infty$ one retrieves
$Y=1/(2\ln 2)$ in agreement with the result for
the fully magnetized case~\cite{AnnSol06}.
We show the two scaling functions together in Fig.~\ref{intercept},
for the example $\delta=10^{-3}$. As expected the two functions agree
in the intermediate regime $1\ll u_t\ll 1/\delta$, where the axis
ratio crosses over from values typical of unmagnetized coarsening
($Y\sim \delta$) to the values of order unity for the magnetized
scenario.
%

\section{Discussion}
\label{sec:crossover_summary}

We have used exact calculations to study the crossover from
unmagnetized to magnetized initial 
conditions in the critical coarsening of the spherical
ferromagnet. Our focus was on the correlation and response functions of
the overall magnetization and the associated non-equilibrium
fluctuation-dissipation (FD) relations. We derived, in particular, the
first exact results (in the nontrivial regime $d<4$) for the crossover
function $X^\infty(u)$ governing 
the behaviour of the asymptotic FD ratio $X^\infty$; $u=\tw/\taum$ is
the appropriate scaling variable, namely the ratio of the earlier measurement
time $\tw$ and the timescale $\taum\sim 1/m_0^2$ set by the initial
magnetization $m_0$. While $X^\infty(u)$ does interpolate between the
known unmagnetized ($u\ll 1$) and fully magnetized ($u\gg 1$) limits, we
found that unexpectedly the behaviour for intermediate $u$ is not
monotonic. In fact, for dimensions $d\approx 2$ close to the lower critical
dimension these non-monotonicities turn into pole singularities in
$X^\infty(u)$.

We traced this unusual behaviour to a non-monotonic dependence on the
earlier time $\tw$ of the two-time magnetization correlator
$C(t,\tw)$, which displays a weak maximum at $u=\order(1)$ and a
corresponding mininum at $u=\order(\delta^{-1/2})$ (for sufficiently
large $u_t=t/\tau_m=\order(\delta^{-1})$). We interpreted the
maximum as the result of an unusually large variance of the
magnetization fluctuations in this region, corresponding to a strong
peak in the {\em equal-time correlator} $C(\tw,\tw)$. The maximum and
minimum of $C(t,\tw)$ also manifest themselves as S-shapes in the
magnetization FD plots for $d\approx 2$.

As an aside, we notice that non-monotonicities in the asymptotic FDR
have previously been observed as a function of the {\em lengthscale}
being probed~\cite{CalGam07}. Also here the effect gets stronger as
$d\to 2$. On the other hand, the non-monotonic dependence on the
lengthscale disappears for large enough $u$ at any $d>2$, so that it
is unclear whether the physical mechanism at work here is related to
the one causing the
complicated dependence of $X^\infty$ on scaled system age that we
saw above.

The calculations for the more general crossover case revealed a
typographical error in our earlier study of fully magnetized initial
conditions~\cite{AnnSol06}. Having corrected this, the expansion to
first order in $4-d$ of the asymptotic FDR $X^\infty$ for the
magnetized case now {\em agrees} with the result of an RG
calculation for the $O(n\to\infty)$ model~\cite{CalGam07}. This
agreement suggests, non-trivially, that the spherical and
$O(n\to\infty)$ models are closely related even beyond the leading
order Gaussian description of their dynamics. One might then suspect
similar agreement also with the $n$-vector model; to verify this, it
would be desirable to extend existing RG expansions
around $d=2$~\cite{FedTri06} beyond the leading term
$X^\infty=1/2+\order(d-2)$. Note that in comparing the spherical with
the $O(n)$ and $n$-vector models one has to look at the {\em
longitudinal} degrees of freedom in the latter since these are the
ones which---like the magnetization in the spherical case---have
nonzero average. The {\em transverse} fluctuations in coarsening
from a state with finite initial magnetization behave
differently, giving in the $O(n\to\infty)$ model an asymptotic FDR of
$X^\infty=d/(d+2)$~\cite{CalGamKrz06}. One expects that the
transverse fluctuations in the $n$-vector model would give the same
FDR for $n\to\infty$. This is consistent with the first-order
expansion $X^\infty=1/2+(d-2)/8$ calculated in~\cite{FedTri06}.
Intriguingly, even though the spherical model with its single degree
of freedom per lattice site has no direct analogue of transverse
fluctuations, it gives the same FDR $X^\infty=d/(d+2)$ for short-range
observables when the system coarsens from an initially magnetized
state.

In future work, an issue of obvious interest would be to understand
how generic our results are, \ie\ whether similar non-monotoniticies
appear also in true short-range models. Field-theoretic calculations
near $d=4$~\cite{CalGam05} for
\eg\ the $O(n)$ model should in principle be possible, and could be
directly compared to the expansion\eq{Xinf_exp} of our results near
$d=4$. 
Our analysis also suggests that if similar expansions were carried out near
$d=2$~\cite{FedTri06}, very rich behaviour could result.

\appendix
\section{Corrections to Ref.~\cite{AnnSol06}}
\label{ref:corrections}

In this appendix we list the required correction to the 
relevant equations of~\cite{AnnSol06}. 
The source of the error was equation $(8.94)$ of~\cite{AnnSol06}: it should 
be replaced by
\be
\fl \frac{CC(t,\tw)}{CC(t,t)}=\G(t/\tw),
\quad
\G(x)=\left\{
\begin{array}{ll}
{\displaystyle \frac{\int d\omt\, \omt^{(d-6)/2} \sc{C}^2(\omt) e^{-2(x-1)\omt}}
{x\int d\omt\, \omt^{(d-6)/2} \sc{C}^2(\omt)}} & \mbox{for $x\geq 1$} \\
x^{(d-4)/2}\G(1/x) & \mbox{for $x\leq 1$}
\end{array}
\right.
\label{CCt_scaling}
\ee
The old version had an erroneous $x^{(d-6)/2}$ rather than $x^{(d-4)/2}$
in the second line of the curly bracket. 
All other mistakes are due to trivial propagation of the one above. 
This affects the first integral in each of equations $(8.99)$ and $(8.100)$, 
whose correct versions are
\bea 
\fl \frac{2\mu_d U}{\CCtd} &=&
\int_0^1\!\! du\, \sc{M}(x/u\yw)u^{(d-2)/4} {\mathcal{G}}(1/u)+
\int_1^{x/\yw}\!\! du\, \sc{M}(x/u\yw)u^{(6-d)/4} {\mathcal{G}}(u)
\\
\fl &=& \int_1^\infty\!\! du\, \sc{M}(xu/\yw)u^{-(d+6)/4} {\mathcal{G}}(u)
+
\int_1^{x/\yw}\!\! du\, \sc{M}(x/u\yw)u^{(6-d)/4} {\mathcal{G}}(u)
\label{decomposition}
\eea
This leads to
%
\bea 
\fl
U&=& \frac{\Tc}{\Gamma(\frac{d-2}{2}) \Gamma(\frac{4-d}{2})}
\left[\int_1^{\infty}\!\!du\, \sc{M}\!\left({\textstyle \frac{xu}{\yw}}\right)u^{-(d+10)/4}
\int_0^1 \!dy \int_0^1 \!dy'\,(yy')^{(d-2)/2}(1-y-y'+u)^{-d/2}\right.
\nn
\fl &&{}+{}\left. \int_1^{x/\yw}\!\!du\, \sc{M}\!\left({\textstyle \frac{x}{u\yw}}\right)u^{(2-d)/4}
\int_0^1 \!dy \int_0^1 \!dy'\, \ldots\right]
\label{exact_I}
\eea
in place of equation $(8.104)$. 
Equations (8.105,107,108,113) as written in terms of $V_d$ are
correct, but $V_d$ itself as stated in $(8.106)$
is incorrect. The correct version is\eq{Vd} in the main text.
The limiting value $V_4$ for $d\to 4$ can be worked out 
explicitly as $V_4=5/6$ and must replace equation $(8.115)$. This 
enters $X^{\infty}$ (denoted $X_{\rm m}^\infty$ in \cite{AnnSol06}) 
only at the first order in an $\epsilon=4-d$-expansion 
\be
X^{\infty}=\frac{4}{5}-\frac{19}{450}\epsilon+\order(\epsilon^2)
\label{X_m_epsilon_expansion}
\ee
which needs to replace equation $(8.116)$.

In the opposite limit $d\to 2$, the error affects again only subleading 
contributions, so \eg\ equation $(8.117)$ stands as written.
The first correction $a_0$ in the 
$\delta=(d-2)/2$-expansion of $V_d=1/\delta+a_0+\ldots$ can be obtained
as $a_0=-3/2$ (rather than $a_0=-1/2-\pi^2/12$~\cite{AnnSol06})
following the reasoning in~\cite{AnnSol06}.
The correct expansion of $X^\infty$ near $d=2$ then becomes
\be
X^\infty=
\frac{1}{2}+\frac{5}{16}(d-2)
+\order((d-2)^2)
\label{X_m_epsilon_prime_expansion}
\ee
which should replace equation $(8.118)$. 

\begin{figure}
\setlength{\unitlength}{0.40mm} 
\begin{picture}(200,145)(-100,10)
\put(20,10){\includegraphics[width=180\unitlength]{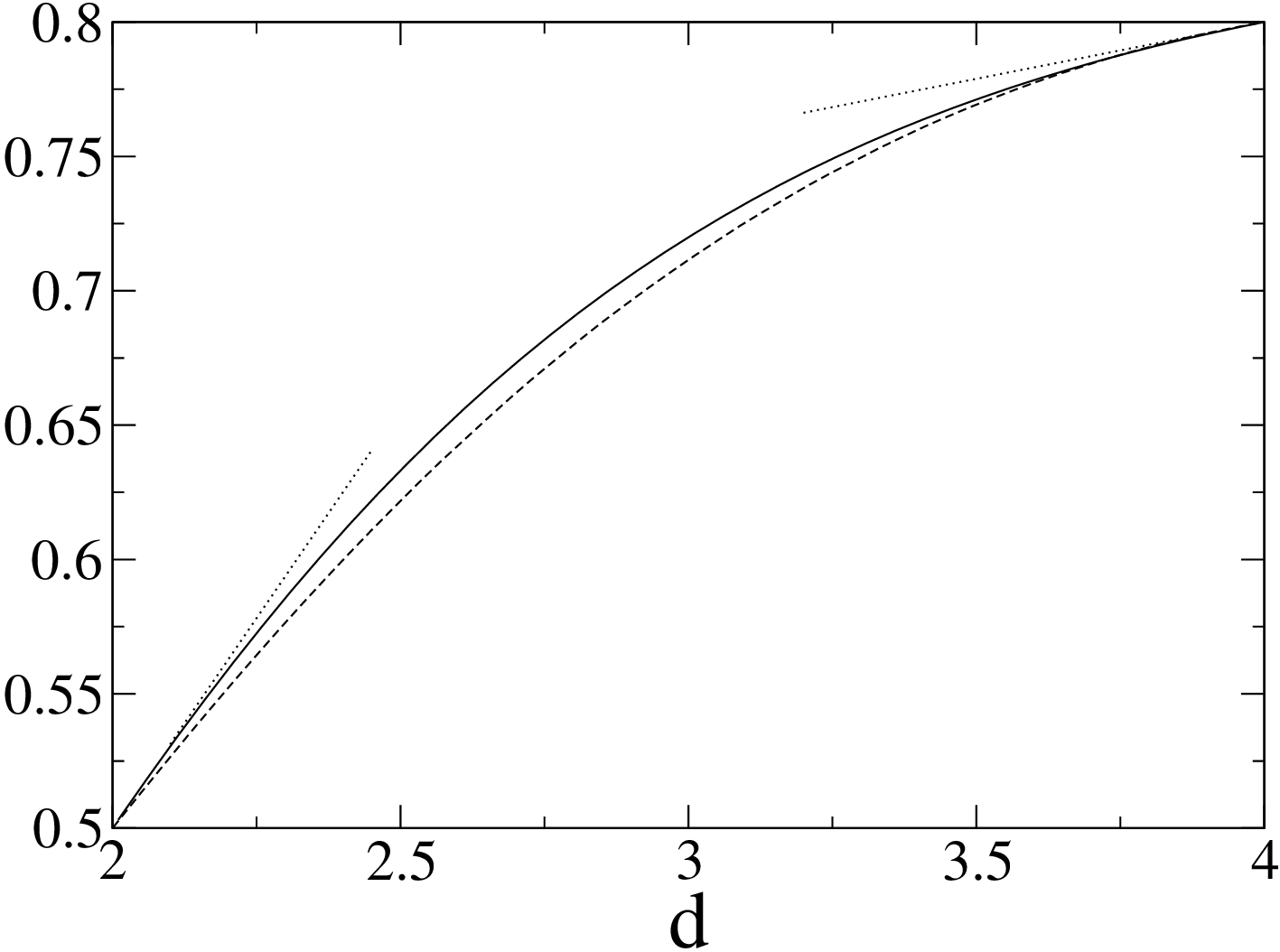}}
\put(0,100){$X^{\infty}$}
\end{picture}
\caption{Asymptotic FDR $X^\infty$ for the magnetization, for
critical coarsening with nonzero initial magnetization. 
Dashed line: old, incorrect version from~\cite{AnnSol06};
solid line: correct version.
dotted lines indicate the (corrected) first-order
expansions\eq{X_m_epsilon_expansion}
and\eq{X_m_epsilon_prime_expansion} near $d=4$ and 
$d=2$, respectively.}
\label{fig:X_magnetized}
\end{figure}
\begin{figure}
\begin{picture}(200,157)(-120,25)
\put(20,40){\includegraphics[width=190\unitlength]{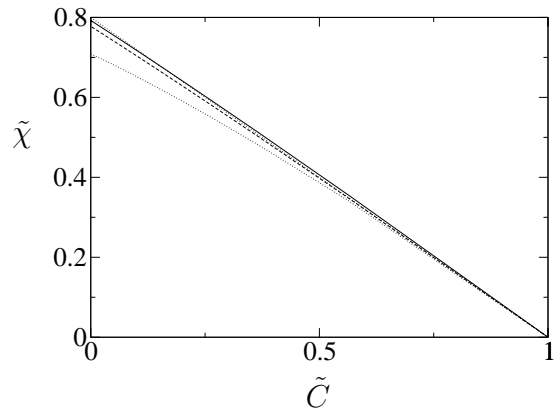}}
\put(5,123){$\tilde{\chi}$}
\put(116,23){$\tilde{C}$}
\end{picture}
\caption{Normalized magnetization FD plot 
showing normalized susceptibility $\tilde{\chi}$ versus normalized
correlation $\tilde{C}$ in the limit of long times. 
For $d=2$ (lower dotted line) and $d=4$ (upper dotted line) the old and the corrected versions coincide.
The deviations between the two versions (correct: full line, old:
dashed line) are largest in $d=3$. The correct plot here is somewhat
closer to the straight line 
obtained for $d=4$, lying very slightly above it in the right-hand part of the plot.
\label{fig:mFD_plot}
}
\end{figure}

Fig.~\ref{fig:X_magnetized} shows the correct $d$-dependence of
$X^\infty$ compared to the erroneous version from~\cite{AnnSol06}. As
expected from the discussion above, the quantitative corrections are
largest around $d=3$ and vanish as $d$ approaches 2 or 4.
For the sake of comparison, we also show in Fig.~\ref{fig:mFD_plot} 
the corrected magnetization FD plots: in $d=2$ and $d=4$ these are as
before, whereas for $d=3$ small quantitative differences are just
about visible.
%
\newpage

\section{Comparison of $\Ltwo$ with previous results}
\label{sec:Ltwo_comparison}

We show briefly in this appendix that the general and exact solution
for $\Ltwo$
\bea
\fl L^{(2)}(t,\tw)&=&-\partial_t N(t,\tw)=
\frac{g(\tw)}{g(t)}\Ltwo\eql(t-\tw)+\frac{g'(t)g(\tw)}{g^2(t)}N\eql(t-\tw)
\label{app:Lexact}
\eea
reproduces the long-time results obtained for the
unmagnetized and fully magnetized limits in~\cite{AnnSol06}. 
Beginning with $d<4$, because $g'(t)/g(t)\sim 1/t$ the second term on
the r.h.s.\ is non-negligible only in the aging regime
($t/\tw=x>1$, hence $t-\tw\sim\tw\gg 1$) where
$N\eql(t-\tw)=[2(t-\tw)/(4-d)]\Ltwo\eql(t-\tw)$
from\eq{Neql}. Inserting\eq{gt_taum} then gives
\bea
\fl L^{(2)}(t,\tw)&=&
L\eql^{(2)}(t-\tw) x^\kappa\left[\frac{u+1}{ux+1}
+\frac{(\alpha ux-\kappa)
(u+1)}{(ux+1)^2}\frac{2}{4-d}\frac{x-1}{x}\right]
\label{Ltwo_cross_below4}
\eea
This result is of the general scaling form
$L^{(2)}(t,\tw)=L\eql^{(2)}(t-\tw)\sc{L}(x,u)$, with the scaling
function $\sc{L}$ providing a multiplicative aging correction of
the equilibrium result.  For $u\ll 1$ or $u\gg 1$ the $u$-dependence
drops out and we obtain the scaling functions found previously
for $d<4$. Specifically, for $u\ll 1$ the square bracket
simplifies to $1/x$ and one gets the unmagnetized result
$\sc{L}(x)=x^{(2-d)/2}$ as in~\cite{AnnSol06}, 
whereas for $u\gg 1$ one retrieves
the expression for the fully magnetized case,
$\sc{L}(x)=x^{(2-d)/2}[2+(2-d)x]/(4-d)$, also derived in~\cite{AnnSol06}.

For $d>4$ the situation is a little more complicated because for long
times, from\eq{Neql},
\be
N\eql(t-\tw)=1/\mu_d+[2(t-\tw)/(d-4)]\Ltwo\eql(t-\tw)
\label{N_dgt4}
\ee
has a constant part of order unity. This means that \eg\ in the
unmagnetized case, where $g(t)$ approaches a constant and one would
normally drop the term proportional to $g'(t)$ in\eq{Ltwo_cross},
a subleading contribution needs to be retained in 
$g'(t)$. The form of this can be found from the Laplace transform\eq{g_Laplace}
together with\eq{LeqLT}: $g'(t)$ has transform 
of
\bea
s\hat{g}(s)-g(0)&=&[s+2\Tc-\hat\Ltwo\eql(0)-(\hat\Ltwo\eql(s)-\hat\Ltwo\eql(0))]
\nn
&&\times\left[\frac{m_0^2}{s}+(1-m_0^2)\hat{f}(0)+(1-m_0^2)(\hat{f}(s)-\hat{f}(0))
\right]
\label{gprime_smalls}
\eea
Transforming to the time domain gives
\be
\fl g'(t)=m_0^2(2\Tc-\hat\Ltwo\eql(0))
+m_0^2\int_t^{\infty}dt'\,\Ltwo\eql(t')-(1-m_0^2)\hat{f}(0)\Ltwo\eql(t)
\label{gprime}
\ee
Here we have neglected the terms arising from $\hat{f}(s)-\hat{f}(0)$,
which decay as $t^{-d/2}$ or even faster at long times and so will be
irrelevant below.
We can now systematically analyse the order of the various
contributions to\eq{app:Lexact} in the long-time limit, obtained by
fixing $u=\tw/\tau_m$ and $x=t/\tw$ and taking $\tw\to\infty$. 
In the second term of\eq{app:Lexact},
$N\eql(t-\tw)$ scales as $\order(\tw^0)+\order(\tw^{(4-d)/2})$
from\eq{N_dgt4}.
For $g'(t)$, we note that in $d>4$ the first bracket in\eq{gprime} is
equal to $\mu_d^{-1}$. Using also 
$m_0^2=c/\taum=cu/\tw$ and $\hat f(0)=c$ gives for long times 
\be
g'(t)=\frac{c}{\mu_d}\left[\frac{u}{\tw}+
\mu_d\left(\frac{2ux}{d-4}-1\right)\Ltwo\eql(t)\right]
\label{gprime2}
\ee
\ie\ $g'(t)=\order(u\tw^{-1})+\order(\tw^{(2-d)/2})$. Integrating
w.r.t.\ $t$ yields $g(t)=\order(\tw^0)+\order(\tw^{(4-d)/2})$;
the leading order term is given explicitly in\eq{gt}. Finally, we 
have $\Ltwo\eql(t-\tw)\sim
(t-\tw)^{(2-d)/2}=\tw^{(2-d)/2}(x-1)^{(2-d)/2}=\order(\tw^{(2-d)/2})$.
One now sees that the unmagnetized case
$u=0$ is special: both terms of\eq{app:Lexact} then scale as
$\order(\tw^{(2-d)/2})$, with $N\eql(t-\tw)=1/\mu_d$,
$g(t)=g(\tw)=c/\mu_d$, $g'(t)=-c\Ltwo\eql(t)$ to leading order so that
\be
\fl
\Ltwo\eql(t,\tw) = 
\Ltwo\eql(t-\tw)-\Ltwo\eql(t) = \Ltwo\eql(t-\tw)\left[
1-\left(\frac{x-1}{x}\right)^{(d-2)/2}\right]
\ee
Aging effects appear again via the multiplicative correction in the
square brackets, which agrees with the result derived in~\cite{AnnSol06}.

In the magnetized case, the first term of\eq{app:Lexact} is still of
$\order(\tw^{(2-d)/2})$ and given by
$\Ltwo\eql(t-\tw)(u+1)/(ux+1)$. The second term, on the other hand, has
a leading $\order(\tw^{-1})$ contribution of
$(1/\mu_d)(u/\tw)(u+1)/(ux+1)^2= (1/\mu_d t)ux(u+1)/(ux+1)^2$. The
subleading terms in $N\eql$, $g$ and $g'$ all give corrections to this
of relative order $\tw^{(4-d)/2}$ which compete with the first term
of\eq{app:Lexact}. The overall result can be written in the form
\bea
\Ltwo(t,\tw)&=&
\Ltwo\eql(t-\tw)\sc{L}(x,u) 
+\frac{1}{\mu_d t}
\frac{ux (u+1)}{(ux+1)^2}
\label{Ltwo_cross_above4}
\eea
In the aging regime ($x>1$) the second term dominates; for
$u\to\infty$ it reduces to $1/(\mu_d t x)=\tw/(\mu_d t^2)$ consistent
with the result of~\cite{AnnSol06}. The full expression for the aging
correction factor $\sc{L}(x,u)$ in the first term is rather long so we
omit it.
%
At any rate, one sees that this first term becomes subleading compared
to the second one already for time differences $t-\tw\sim [\tw
(u+1)/u]^{2/(d-2)} \ll \tw$ where $\sc{L}(x,u)=\sc{L}(1,u)=1$. The
detailed form of the multiplicative aging correction therefore never becomes relevant.

\newpage

\bibliographystyle{unsrt}
\bibliography{references_thesis1109}

\end{document}